\documentstyle[12pt,epsf]{article}
\begin{document}
\hfill{DESY 99-053}

\vspace*{3cm}
\begin{center}
{\large \bf Semileptonic $B_c$-meson decays in sum rules of QCD and NRQCD.}\\
\vspace*{5mm}
V.V. Kiselev$^{a)}$, A.K.~Likhoded$^{a)}$, A.I. Onishchenko$^{b)}$\\
\end{center}
\begin{center}
$^{a)}${\it State Research Center of Russia 
"Institute for High Energy Physics",\\ Protvino, Moscow region, 
142284 Russia}\\
\end{center}
\begin{center}
$^{b)}${\it Institute for Theoretical and Experimental Physics,\\ 
Moscow, 117218 Russia}.
\end{center}

\begin{abstract}
The semileptonic $B_c$-meson decays into the heavy
quarkonia $J/\psi (\eta_c)$ and a pair of leptons are investigated
in the framework of 
three-point sum rules of QCD and NRQCD. Calculations of analytical expressions
for the spectral densities of QCD and NRQCD correlators with account for the
Coulomb-like $\alpha_s/v$ terms are presented. The generalized relations due 
to the spin symmetry of NRQCD for the form factors of $B_c\to J/\psi 
(\eta_c)l\nu_l$ transitions with $l$ denoting one of the leptons $e, \mu $ or
$\tau $,  are derived at the recoil momentum close to zero. 
This allows one to 
express all NRQCD form factors through a single universal quantity, an 
analogue of Isgur-Wise function at the maximal invariant mass of lepton 
pair. The gluon 
condensate corrections to three-point functions are calculated both in full QCD
in the Borel transform scheme and in NRQCD in the moment scheme. This 
enlarges the parameteric stability region of sum rule method, that makes the
results of the approach to be more reliable. Numerical estimates of widths for
the transitions of $B_c\to J/\psi (\eta_c)l\nu_l$ are presented.
\end{abstract}

\newpage
\section{Introduction}

Recently, the CDF collaboration reported on the first experimental observation
of the $B_c$ meson, the heavy quarkonium with the mixed heavy flavour
\cite{cdf}. This meson stands among the families of charmonium $\bar c c$ and
bottominum $\bar b b$ in what concerns its spectroscopic properties: two heavy
quarks move nonrelativistically, since the confinement scale, determining the
presence of light degrees of freedom (sea of gluons and quarks), is suppressed
with respect to the heavy quark masses $m_Q$ as well as the Coulomb-like
exchanges result in transfers about $\alpha_s m_Q^2$, which is again much less
than the heavy quark mass. So, the nonrelativistic picture of binding the
quarks leads to the well-known arrangement of system levels, which is very
similar for the families mentioned above. The calculations of $\bar b c$-mass
spectrum were reviewed in \cite{eichten,prd1}. So, we expect
$$
M_{B_c} = 6.25\pm 0.03\; {\rm GeV,}
$$
when the measurement gave
$$
M^{exp}_{B_c} = 6.40\pm 0.39\pm 0.13\; {\rm GeV.}
$$

There is an essential  difference in the production mechanisms of heavy
quarkonia $\bar c c$, $\bar b b$ and $B_c$. To bind the $\bar b$ and $c$
quarks, one has to produce four heavy quarks in the flavour conserving
interactions\footnote{We do not consider for the production in weak
interactions.}, which allows one to use the perturbative QCD, since the
virtualities are determined by the scale of heavy quarks mass. Thus, we see
that the production of $B_c$ is relatively suppressed $\sigma(B_c)/\sigma(\bar
b b) \sim 10^{-3}$ because of the additional heavy quark pair in final states.
The basic peculiarities of production mechanisms appear due to the higher
orders of QCD even in the leading approximation: the fragmentation regime at
high transverse momenta much greater than the quark masses and a strong role of
nonfragmentational contributions at $p_T\sim m_Q$, which can be exactly
calculated perturbatively; a negligible contribution of octet mechanism
\cite{octet} because there is no enhancement due to a lower order in
$\alpha_s$. The predictions for the  
cross sections and distributions of $B_c$ in
various interactions are discussed in \cite{prod}, where we see a good
agreement with the meausrements of CDF [1]. 

 In contrast to $\bar c c$
and $\bar b b$ states decaying due to the annihilation into the light quarks
and gluons, the $B_c$ meson is a long 
lived particle, since it decays due to the
weak interaction. The lifetime and various modes of decays were analyzed in the
framework of a) potential models \cite{pm}, b) technique of Operator Product
Expansion in the effective theory of NRQCD \cite{nrqcd}, considering the series
in both a small relative velocity $v$ of heavy quarks inside the meson and the
inverse heavy quark mass \cite{beneke}, c) QCD sum rules \cite{Shif,Rein}
applied to the three-point correlators \cite{Col,bagan,Kis2}. The results of
potential models and NRQCD are in agreement with each other. So, we expect
that the total lifetime is equal to
$$
\tau_{B_c} = 0.55\pm 0.15\; {\rm ps,}
$$
which agrees with the experimental value given by CDF [1]:
$$
\tau^{exp}_{B_c} = 0.46^{+0.18}_{-0.16}\pm 0.03\; {\rm ps,}
$$
within the accuracy available. 

Further, the consideration of exclusive $B_c$ decays in the framework of QCD
sum rules indicated that the role of Coulomb corrections to the bare quark loop
results could be very important to reach the agreement with the other
approaches mentioned \cite{Kis2}. This requires working out 
the $\alpha_s/v$
corrections in NRQCD, which possesses a spin symmetry providing some relations
between the exclusive form factors. For the semileptonic decays, such a
relation was derived in \cite{Jenkins}. Note, that the CDF Collaboration
observed 20 events of $B_c^+\to \psi e^+(\mu^+) \nu$, so that the consistent
calculation of semileptonic decay modes is of interest. For theoretical
reviews on the $B_c$ meson physics, see \cite{review}.

In this paper we perform a detailed analysis of semileptonic $B_c$ decays in
the
framework of sum rules in QCD and NRQCD. We recalculate the double spectral
densities available previously in \cite{Col} in full QCD for the massless
leptons and add the analytical expressions for the form factors necessary in
evaluation of decays to massive leptons and P-wave levels of quarkonium with
different quark masses. We analyze the NRQCD sum rules for the three-point
correlators for the first time. We derive generalized relations between the
NRQCD form factors, which extends the consideration in \cite{Jenkins}, because
we explore a soft limit of recoil momentum close to zero, wherein the
velocities of initial and final heavy quarkonia $v_{1,2}$ are not equal to each
other, when their product tends to zero, in contrast to the hard limit
$v_1=v_2$. The spin symmetry relations between the form factors are conserved
after the taking into account the Coulomb $\alpha_s/v$ corrections, which can
be written down in covariant form. We investigate numerical estimates in the
sum rules schemes of spectral density moments and Borel transform and show an
important role of Coulomb corrections. Next, we perform the calculation of
gluon condensate contribution to the three-point sum rules of both full QCD and
NRQCD for the case of three massive quarks, for the first time.

The paper is organized as follows. The QCD sum rules of three-point correlators
are considered in Section 2, where the spectral densities are calculated in the
bare quark-loop approximation and with account of the Coulomb corrections, and
the gluon condensate term in the Borel transform scheme is presented. 
Numerical estimates of semileptonic decay modes are also given here.
 Section 3 is devoted
to the NRQCD sum rules for recoil close to zero. The spin symmetry relations
are
derived and the gluon condensate is taken into account in the scheme of
moments. The results are summarized in Section 4. Appendices A and B contain
technical details in evaluation of decay widths for the massive leptons and
gluon condensate in full QCD, respectively.

\section{Three-point QCD sum rules}

In this paper we will use the approach of three-point QCD sum rules
\cite{Shif,Rein} in the study of form factors and decay rates for the
transitions $B_c^{+}\to \psi (\eta_c)l^+\nu_l$, where $l$ denotes one of the
leptons $e, \mu$ or $\tau$. This procedure is similar to that of  two-point
sum rules and the information from the latter on the coupling of mesons to
their currents is required in order to extract the values for the form factors.
Thus, in our work we will use the meson couplings, defined by the following
equations:
\begin{equation}
\langle 0|\bar q_1 i\gamma_5 q_2|P(p)\rangle = \frac{f_{P}M_{P}^2}{m_1 + m_2},
\end{equation}
and
\begin{equation}
\langle 0|\bar q_1\gamma_{\mu} q_2|V(p,\epsilon)\rangle =
i\epsilon_{\mu}M_Vf_V,
\end{equation}
where $P$ and $V$ represent the scalar and vector mesons with desired flavour
quantum numbers, respectively, and $m_1, m_2$ are the quark masses. Now we
would like to
describe the method used.

\subsection{Description of the method}

As we have already said, for the calculation of hadronic matrix elements
relevant to the semileptonic $B_c$-decays into the pseudoscalar and vector
mesons in the framework of QCD, 
we explore the QCD sum rule method. The hadronic
matrix elements for the transition $B_c^{+}\to \psi (\eta_c)l^+\nu_l$ can be
written down as follows:
\begin{eqnarray}
\langle\eta_c(p_2)|V_{\mu}|B_c(p_1)\rangle &=& f_{+}(p_1 + p_2)_{\mu} +
f_{-}q_{\mu},\\
\frac{1}{i}\langle J/\psi (p_2)|V_{\mu}|B_c(p_1)\rangle &=& 
i F_V\epsilon_{\mu\nu\alpha\beta}\epsilon^{*\nu}(p_1 +
p_2)^{\alpha}q^{\beta},\\
\frac{1}{i}\langle J/\psi (p_2)|A_{\mu}|B_c(p_1)\rangle &=&
F_0^A\epsilon_{\mu}^{*} + 
F_{+}^{A}(\epsilon^{*}\cdot p_1)(p_1 + p_2)_{\mu} + 
F_{-}^{A}(\epsilon^{*}\cdot p_1)q_{\mu}, 
\end{eqnarray}
where $q_{\mu} = (p_1 - p_2)_{\mu}$ and $\epsilon^{\mu} = \epsilon^{\mu}(p_2)$
is the polarization vector of $J/\psi$-meson. $V_{\mu}$ and $A_{\mu}$ are the
flavour changing vector and axial electroweak currents. The form factors
$f_{\pm}, F_V, F_0^A$ and $F_{\pm}^A$ are functions of $q^2$ only. It should be
noted, that by virtue of transversality of the lepton current $l_{\mu} =
l\gamma_{\mu}(1 + \gamma_5)\nu_l$ in the limit $m_l\to 0$, the probabilities of
semileptonic decays into $e^{+}\nu_e$ and $\mu^{+}\nu_{\mu}$ are independent of
$f_{-}$ and $F_{-}^A$. Thus, in calculation of these particular decay modes of
$B_c$-meson these form factors can be consistently neglected
\cite{Col,pm,Kis2}. However, since the calculation of both the semileptonic
decay modes, including $e,\mu$ or $\tau$, and some  hadronic decays, stands
among the goals of this paper, we will present the results for the complete set
of form factors given in Eqs. (3)-(5). 

Following the standard procedure for the evaluation of form factors in the
framework of QCD sum rules, we consider the three-point functions:
\begin{eqnarray}
\Pi_{\mu}(p_1, p_2, q^2) &=& i^2 \int dxdye^{i(p_2\cdot x - p_1\cdot
y)} \cdot \nonumber\\
&&\langle 0|T\{\bar q_1(x)\gamma_5 q_2(x), V_{\mu}(0), \bar b(y)\gamma_5
c(y)\}|0
\rangle,\\
\Pi_{\mu\nu}^{V, A}(p_1, p_2, q^2) &=& i^2 \int dxdye^{i(p_2\cdot x - p_1\cdot
y)} \cdot \nonumber\\
&&  \langle 0|T\{\bar q_1(x)\gamma_{\mu} q_2(x), J_{\mu}^{V, A}(0), 
\bar b(y)\gamma_5 c(y)\}|0\rangle,
\end{eqnarray}
where $\bar q_1(x)\gamma_5 q_2(x)$ and $\bar q_1(x)\gamma_{\nu}q_2(x)$ are
interpolating
currents for states with the quantum numbers of $\eta_c$ and $J/\psi$,
correspondingly. 
$J_{\mu}^{V, A}$ are the currents $V_{\mu}$ and $A_{\mu}$ of relevance to the
various cases.

The Lorentz structures in the correlators can be written down as:
\begin{eqnarray}
\Pi_{\mu} &=& \Pi_{+}(p_1 + p_2)_{\mu} + \Pi_{-}q_{\mu},\label{R1}\\
\Pi_{\mu\nu}^V &=& i\Pi_V\epsilon_{\mu\nu\alpha\beta}p_2^{\alpha}p_1^{\beta},\\
\Pi_{\mu\nu}^A &=& \Pi_{0}^{A}g_{\mu\nu} + \Pi_{1}^{A}p_{2, \mu}p_{1, \nu} +
\Pi_{2}^{A}p_{1, \mu}p_{1, \nu} + \Pi_{3}^{A}p_{2, \mu}p_{2, \nu} + 
\Pi_{4}^{A}p_{1, \mu}p_{2, \nu}.
\end{eqnarray}
\noindent
The form factors $f_{\pm}$, $f_V$, $F_{0}^{A}$ and $F_{\pm}^{A}$ will be
determined, respectively, from the amplitudes $\Pi_{\pm}$, $\Pi_V$,
$\Pi_{0}^{A}$ and $\Pi_{\pm}^{A} = \frac{1}{2}(\Pi_{1}^{A}\pm \Pi_{2}^{A})$. In
(8)-(10) the scalar amplitudes $\Pi_i$ are the functions of kinematical
invariants, i.e. $\Pi_i = \Pi_i(p_1^2, p_2^2, q^2)$. 

To calculate the QCD expression for the three-point correlators we employ the
Operator Product Expansion (OPE) for the $T$-ordered product of currents in
(6)-(7). The vacuum correlations of heavy quarks are related to their
contribution to the gluon operators. For example, for the $\langle\bar
QQ\rangle$ and $\langle\bar QGQ\rangle$ condensates the heavy quark expansion
gives
\begin{eqnarray}
\langle\bar QQ\rangle &=& -\frac{1}{12m_Q}\frac{\alpha_s}{\pi}\langle
G^2\rangle -
\frac{1}{360m_Q^3}\frac{\alpha_s}{\pi}\langle G^3\rangle + ...\nonumber\\
\langle\bar QGQ\rangle &=& \frac{m_Q}{2}\log (m_Q^2)\frac{\alpha_s}{\pi}\langle 
G^2\rangle - \frac{1}{12m_Q}\frac{\alpha_s}{\pi}\langle G^3\rangle + ...
\nonumber
\end{eqnarray}
Then, in the lowest order for the energy dimension of operators the only
nonperturbative correction comes from the gluon condensate:
\begin{equation}
\Pi_i(p_1^2, p_2^2, q^2) = \Pi_i^{pert}(p_1^2, p_2^2, q^2) + 
\Pi_i^{G^2}(p_1^2, p_2^2, q^2)\langle\frac{\alpha_s}{\pi}G^2\rangle .
\label{OPE}
\end{equation}
The leading QCD term is a triangle quark loop diagram, for which we can write
down the double dispersion representation at $q^2\leq 0$:
\begin{equation}
\Pi_i^{pert}(p_1^2, p_2^2, q^2) = -\frac{1}{(2\pi)^2}\int
\frac{\rho_i^{pert}(s_1, s_2, Q^2)}{(s_1 - p_1^2)(s_2 - p_2^2)}ds_1ds_2 + 
\mbox{subtractions},
\label{pertdisp}
\end{equation}
where $Q^2 = -q^2 \geq 0$. The integration region in (\ref{pertdisp}) is
determined by the condition
\begin{equation}
-1 < \frac{2s_1s_2 + (s_1 + s_2 - q^2)(m_b^2 - m_c^2 - s_1)}
{\lambda^{1/2}(s_1, s_2, q^2)\lambda^{1/2}(m_c^2, s_1, m_b^2)} < 1,
\end{equation}
and $$\lambda(x_1, x_2, x_3) = (x_1 + x_2 - x_3)^2 - 4x_1x_2.$$ The calculation
of spectral densities $\rho_i^{pert}(s_1, s_2, Q^2)$ and gluon condensate
contribution to (\ref{OPE}) will be considered in underlying sections. Now let
us proceed with the physical part of three-point sum rules. The connection to
hadrons in the framework of QCD sum rules is obtained by matching the resulting
QCD expressions of current correlators with the spectral representation,
derived from a double dispersion relation at $q^2\leq 0$.
\begin{equation}
\Pi_i(p_1^2, p_2^2, q^2) = -\frac{1}{(2\pi)^2}\int
\frac{\rho_i^{phys}(s_1, s_2, Q^2)}{(s_1 - p_1^2)(s_2 - p_2^2)}ds_1ds_2 + 
\mbox{subtractions}.
\label{physdisp}
\end{equation}
Assuming that the dispersion relation (\ref{physdisp}) is well convergent, the
physical spectral functions are generally saturated by the lowest lying
hadronic states plus a continuum starting at some effective thresholds
$s_1^{th}$ and $s_2^{th}$:
\begin{eqnarray}
\rho_i^{phys}(s_1, s_2, Q^2) &=& \rho_i^{res}(s_1, s_2, Q^2) + \\
&& \theta (s_1-s_1^{th})\cdot\theta (s_2-s_2^{th})\cdot
\rho_i^{cont}(s_1, s_2, Q^2),
\nonumber
\end{eqnarray}
where
\begin{eqnarray}
\rho_i^{res}(s_1, s_2, Q^2) &=& {\langle 0|\bar c\gamma_{\mu}(\gamma_5)
c|J/\psi (\eta_c) \rangle\langle J/\psi (\eta_c)|F_i(Q^2)|B_c\rangle\langle
B_c|\bar b\gamma_5 c|0)\rangle}\cdot \nonumber\\
&& {(2\pi)^2 \delta(s_1-M_1^2) \delta(s_2-M_2^2)}
+ \mbox{higher~state~contributions},
\end{eqnarray}
where $M_{1,2}$ denote the masses of quarkonia in the initial and final states.
The continuum of higher states is modelled by the perturbative absorptive part
of $\Pi_i$, i.e. by $\rho_i$.  Then, the expressions for the form factors $F_i$
can be derived by equating the representations for the three-point functions
$\Pi_i$ in (\ref{OPE}) and (\ref{physdisp}), which means the formulation of sum
rules.

\subsection{Calculating the spectral densities}

In this section we present  the analytical expressions to one loop
approximation for the perturbative spectral functions. We have recalculated
their values, already available in the literature \cite{Col}. Among new results
there are the expressions for $\rho_{-}, \rho_{-}^{A}$ and $\rho_{\pm}^{'A}$,
where $\rho_{\pm}^{'A}$ are spectral functions, which come from the double
dispersion representation of $\Pi_{\pm}^{'A} = \frac{1}{2}(\Pi_{3}^{A}\pm
\Pi_{4}^{A})$. These spectral densities are not required for the purposes of
this paper, but they will be useful for calculation of form factors for the
transition of $B_c$-meson into a scalar meson\footnote{ The meson at the
$P$-wave level, for which $\langle 0|\bar q_1\gamma_{\mu}q_2|P(p)\rangle = i
f_Pp_{\mu},$ where $P(p)$ denotes the scalar $P$-wave meson under
consideration, and $m_1\neq m_2$.}. The procedure of evaluating the spectral
functions involves the standard use of Cutkosky rules \cite{Cutk}. There is,
however, one subtle point in using these rules. At $Q^2 < 0$ there is no
problem in applying the Cutkosky rules in order to determine $\rho_i(s_1, s_2,
Q^2)$ and the limits of integration over $s_1, s_2$. At $Q^2 >0$, which is the
physical region, non-Landau-type singularities appear \cite{Ball1,Ball2}, what
makes the determination of spectral functions to be quite complicated. In our
case we restrict the region of integration in $s_1$ and $s_2$ by $s_1^{th}$ and
$s_2^{th}$, so that at moderate values of $Q^2$ the non-Landau singularities do
not contribute to the values of spectral functions. For spectral densities
$\rho_i(s_1, s_2, Q^2)$ we have the following expressions:
\begin{eqnarray}
\rho_{+}(s_1, s_2, Q^2) &=& \frac{3}{2k^{3/2}}\{\frac{k}{2}(\Delta_1 +
\Delta_2) -
k[m_3(m_3 - m_1) + m_3(m_3 - m_2)] - \nonumber\\
&& [2(s_2\Delta_1 + s_1\Delta_2) - u(\Delta_1 + \Delta_2)]\\
&& \cdot [m_3^2 - \frac{u}{2} + m_1m_2 - m_2m_3 - m_1m_3]\}, \nonumber \\
\rho_{-}(s_1, s_2, Q^2) &=& - \frac{3}{2k^{3/2}}\{\frac{k}{2}(\Delta_1 -
\Delta_2) -
k[m_3(m_3 - m_1) - m_3(m_3 - m_2)] + \nonumber\\
&& [2(s_2\Delta_1 - s_1\Delta_2) + u(\Delta_1 - \Delta_2)]\\
&& \cdot [m_3^2 - \frac{u}{2} + m_1m_2 - m_2m_3 - m_1m_3]\}, \nonumber \\
\rho_{V}(s_1, s_2, Q^2) &=& \frac{3}{k^{3/2}}\{(2s_1\Delta_2 - u\Delta_1)(m_3 -
m_2)
\nonumber \\
&& + (2s_2\Delta_1 - u\Delta_2)(m_3 - m_1) + m_3k\}, \\
\rho_{0}^A(s_1, s_2, Q^2) &=& \frac{3}{k^{1/2}}\{
(m_1 - m_3)[m_3^2 + \frac{1}{k}(s_1\Delta_2^2 + s_2\Delta_1^2 -
u\Delta_1\Delta_2)]
\nonumber \\
&& - m_2(m_3^2 - \frac{\Delta_1}{2}) - m_1(m_3^2 - \frac{\Delta_2}{2}) \\
&& + m_3[ m_3^2 - \frac{1}{2}(\Delta_1 + \Delta_2 - u) + m_1m_2]\}, \nonumber\\
\rho_{+}^A(s_1, s_2, Q^2) &=& \frac{3}{2k^{3/2}}\{m_1[2s_2\Delta_1 - u\Delta_1
+
4\Delta_1\Delta_2 + 2\Delta_2^2]\nonumber \\
&& m_1m_3^2[4s_2 - 2u] + m_2[2s_1\Delta_2 - u\Delta_1] - m_3[2(3s_2\Delta_1 +
s_1\Delta_2)
\nonumber \\
&& - u(3\Delta_2 + \Delta_1) + k + 4\Delta_1\Delta_2 + 2\Delta_2^2 + m_3^2(4s_2
- 2u)] \\
&& + \frac{6}{k}(m_1 - m_3)[4s_1s_2\Delta_1\Delta_2 - u(2s_2\Delta_1\Delta_2 + 
s_1\Delta_2^2 + s_2\Delta_1^2)\nonumber \\
&& + 2s_2(s_1\Delta_2^2 + s_2\Delta_1^2)]\}, \nonumber \\
\rho_{-}^A(s_1, s_2, Q^2) &=& -\frac{3}{2k^{5/2}}\{
kum_3(2m_1m_3 - 2m_3^2 + u) + 12(m_1 - m_3)s_2^2\Delta_1^2 + \nonumber \\
&& k\Delta_2[(m_1 + m_3)u - 2s_1(m_2 - m_3)] + 2\Delta_2^2(k + 3us_1)(m_1-m_3) 
\nonumber \\
&& + \Delta_1[ku(m_2 - m_3) + 2\Delta_2(k - 3u^2)(m_1 - m_3)] + \\
&& 2s_2(m_1 - m_3)[2km_3^2 - k\Delta_1 + 3u\Delta_1^2 - 6u\Delta_1\Delta_2] - 
\nonumber\\
&& 2s_1s_2(km_3 - 3\Delta_2^2(m_1 - m_3))],\nonumber \\
\rho_{+}^{'A}(s_1, s_2, Q^2) &=& -\frac{3}{2k^{5/2}}\{ -2(m_1 - m_3)
[(k - 3us_2)\Delta_1^2 + 6s_1^2\Delta_2^2] + \nonumber \\
&& ku(m_1 - m_3)(2m_3^2 + \Delta_2) + ku^2m_3 + \Delta_1[ku(2m_1 - m_2 -
3m_3)\nonumber \\ 
&& - 2(m_1 - m_3)(ks_2 - k\Delta_2 + 3u^2\Delta_2)] - \\
&& 2s_1[(m_1-m_3)(2km_3^2 - 6u\Delta_1\Delta_2 - 3u\Delta_2^2) + \nonumber \\
&& 2s_2(km_3 + 3m_1\Delta_1^2 - 3m_3\Delta_1^2) + k\Delta_2(2m_1 - m_2 -
3m_3)]\},\nonumber \\
\rho_{-}^{'A}(s_1, s_2, Q^2) &=& \frac{3}{2k^{5/2}}\{ 2(m_1 - m_3)
[(k + 3us_2)\Delta_1^2 + 6s_1^2\Delta_2^2] + \nonumber \\
&& ku(m_1 - m_3)(2m_3^2 + \Delta_2) + ku^2m_3 + \Delta_1[ku(- 2m_1 - m_2 +
m_3)\nonumber \\ 
&& - 2(m_1 - m_3)(ks_2 - k\Delta_2 + 3u^2\Delta_2)] + \\
&& 2s_1[(m_1-m_3)(2km_3^2 - 6u\Delta_1\Delta_2 + 3u\Delta_2^2) - \nonumber \\
&& 2s_2(km_3 - 3m_1\Delta_1^2 + 3m_3\Delta_1^2) + k\Delta_2(2m_1 + m_2 -
m_3)]\}.\nonumber
\end{eqnarray}
Here $k = (s_1 + s_2 + Q^2)^2 - 4s_1s_2$, $u = s_1 + s_2 + Q^2$, $\Delta_1 =
s_1 - m_1^2 + m_3^2$ and $\Delta_2 = s_2 - m_2^2 + m_3^2$. $m_1, m_2$ and $m_3$
are the masses of quark flavours relevant to the various decays, see
prescriptions in Fig. 1. 

\setlength{\unitlength}{1mm}

\begin{figure}[th]

\vspace*{-2cm}
\begin{center}
\begin{picture}(80,80)
\put(0,0){\epsfxsize=3cm \epsfbox{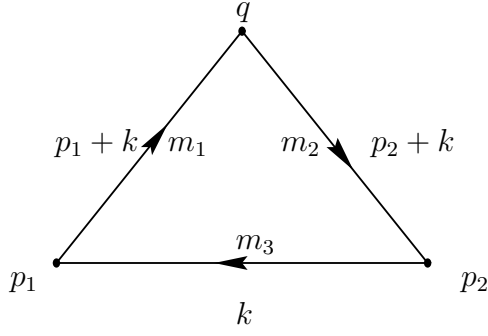}}

\put(0,20){$p_1$}
\put(60,20){$p_2$}
\put(30,15){$k$}
\put(30,56){$q$}
\put(30,25){$m_3$}
\put(6,38){$p_1+k$}
\put(48,38){$p_2+k$}
\put(21,38){$m_1$}
\put(36,38){$m_2$}

\end{picture}
\end{center}

\vspace*{-2cm}
\caption{The triangle diagram, giving the leading perturbative term in the OPE
expansion of three-point function.}
\end{figure}
\normalsize

We neglect hard $O(\frac{\alpha_s}{\pi})$ corrections to the triangle diagrams,
as they are not available yet. Nevertheless, we expect that their contributions
are quite small $\sim 10\%$ and so, taking into account the accuracy of QCD sum
rules, the correction will not change drastically our results.

In expressions (\ref{pertdisp}) the integration over $s_1$ and $s_2$ is
performed in the near-threshold region, where instead of $\alpha_s$, the
expansion should be done in the parameters $(\alpha_s/v_{13(23)})$, with
$v_{13(23)}$ meaning the relative velocities of quarks in $(b\bar c)$ and
$(c\bar c)$ systems. For the heavy quarkonia, where the quark velocities are
small, these corrections take an essential role (as it is the case for
two-point sum rules \cite{Novikov,Schw}). The $\alpha_s/v$ corrections, caused
by the Coulomb-like interaction of quarks, are related with the ladder
diagrams, shown in Fig. 2. It is well known, that an account of these
corrections in two-point sum rules numerically leads to a double-triple
multiplication of Born value of spectral density \cite{fbc,Grigor}.

\setlength{\unitlength}{1.5mm}

\begin{figure}[th]

\vspace*{-6cm}
\begin{center}
\hspace*{1cm}
\begin{picture}(80,80)
\put(0,0){\epsfxsize=9cm \epsfbox{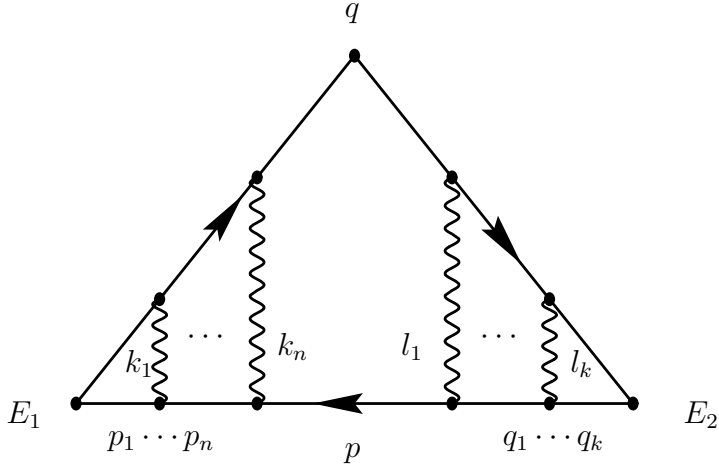}}

\put(0,0){$E_1$}
\put(60,0){$E_2$}
\put(30,-3){$p$}
\put(30,36){$q$}
\put(9,-2){$p_1\cdots p_n$}
\put(44,-2){$q_1\cdots q_k$}
\put(16,7){$\cdots$}
\put(42,7){$\cdots$}
\put(10.5,4.5){$k_1$}
\put(24,6){$k_n$}
\put(35,6){$l_1$}
\put(50,4.5){$l_k$}

\end{picture}
\end{center}

\caption{The ladder diagram  of the Coulomb-like quark interaction.}
\end{figure}

\setlength{\unitlength}{1mm}

\normalsize

Now, let us comment the effect of these corrections in the case of
three-point sum rules \cite{Kis2}. Consider, for example, the three-point
function $\Pi_{\mu}(p_1, p_2, q)$ at $q^2 = q_{max}^2$, where $q_{max}^2$ is
the maximum invariant mass of the lepton pair in the decay $B_c\to\eta_c
l\nu_l$. Introduce the notations $p_1\equiv (m_b + m_c +E_1,\vec 0)$ and
$p_2\equiv (2m_c + E_2,\vec 0)$. At $s_1 = M_1^2$ and $s_2 = M_2^2$ we have
$E_1\ll (m_b + m_c)$ and $E_2\ll 2m_c$. In this kinematics, the quark
velocities are small, and, thus, the diagram in Fig. 2 may be considered in the
nonrelativistic approximation. We will use the Coulomb gauge, in which the
ladder diagrams with the Coulomb-like gluon exchange are dominant. Then the
gluon propagator has the form
\begin{equation}
D^{\mu\nu} = i\delta^{\mu 0}\delta^{\nu 0}/{\bf k}^2.
\end{equation}      
In this approximation, the nonrelativistic potential of heavy quark interaction
in the momentum  representation is given by
\begin{eqnarray}
\tilde V({\bf k}) &=& -\frac{4}{3}\alpha_s({\bf k}^2)\frac{4\pi}{{\bf k}^2},
 \quad 
\alpha_s({\bf k}^2) = \frac{4\pi}{b_0 ln({\bf k}^2/\Lambda^2)}, \nonumber\\
b_0 &=& 11 -\frac{2}{3}n_f, \quad \Lambda = \Lambda_{\overline{MS}}\exp\left [
\frac{1}{b_0}\left (\frac{31}{6} - \frac{5}{9}n_f\right )\right ],\nonumber
\end{eqnarray}
with $n_f$ being the number of flavours, while the fermionic propagators,
corresponding to either a particle or antiparticle, have the following forms:
\begin{eqnarray}
S_F(k + p_i) &=& \frac{i(1 + \gamma_0)/2}{E_i + k^0 - \frac{|{\bf k}|^2}{2m} +
i0},\nonumber\\
S_F(p) &=& \frac{-i(1-\gamma_0)/2}{-k^0 - \frac{|{\bf k}|^2}{2m} +
i0}.\nonumber
\end{eqnarray} 
The notations, concerning Fig. 2, are given by
$$
k_i = p_{i+1} - p_i,\quad l_i = q_i - q_{i-1},\quad p_{n+1}\equiv q_0\equiv p.  
$$   
Integration over $p_i^0, p^0$ and $q_i^0$ by means of residues yields the
following expression
\begin{eqnarray}
\Pi_{\mu}(E_1, E_2, q) &=& 2 N_c g_{\mu 0}\sum_{n=1}^{\infty}
\prod_{i=1}^n \frac{d {\bf p}_i}{(2\pi)^3(\frac{|{\bf p}_i|^2}{2\mu_1} - E_1
-i0)} \tilde V (({\bf p}_{i+1} - {\bf p}_i)^2)\cdot \nonumber\\
&& \frac{1}{\frac{|{\bf p}|^2}{2\mu_1} - E_1 - i0}\cdot
\frac{1}{\frac{|{\bf p}|^2}{2\mu_2} - E_2 - i0}\cdot\label{C1}\\
&& \sum_{k=1}^{\infty}\prod_{j=1}^{k}\frac{d {\bf q}_j}
{(2\pi )^3(\frac{|{\bf q}_j|^2}{2\mu_2} - E_2 - i0)}\tilde V (({\bf q}_j - 
{\bf q}_{j-1})^2)\frac{d {\bf p}}{(2\pi)^3}\nonumber ,
\end{eqnarray}      
$$
\tilde V(({\bf p}_{n+1} - {\bf p}_n)^2)\equiv \tilde V(({\bf p} - {\bf
p}_n)^2), \quad
\tilde V(({\bf q}_{1} - {\bf q}_0)^2) = \tilde V(({\bf q}_{1} - {\bf p})^2),
$$
where $N_c$ denotes the number of colors,
$\mu_1$ and $\mu_2$ are the reduced masses of the $(b\bar c)$ and $(c\bar c)$
systems, correspondingly. This three-point function may be expressed in terms
of the Green's functions for the relative motion of heavy quarks in the $(b\bar
c)$ and $(c\bar c)$ systems in the Coulomb field, $G_E^{(i)}(\bf x, y)$:
\begin{eqnarray}
G_E^{(i)}({\bf x, y}) &=& \sum_{n=1}^{\infty}\left (\prod_{k=1}^n\int
\frac{d {\bf p}_k}{(2\pi )^3(\frac{|{\bf p}_k|^2}{2\mu_i} - E_i - i0)}\right
)\nonumber\\
&& \times\prod_{k=1}^{n-1}\tilde V (({\bf p}_k - {\bf p}_{k+1})^2)
e^{i{\bf p}_1{\bf x} - i{\bf p}_n{\bf y}}. \label{C2}    
\end{eqnarray}
Comparing the expressions (\ref{C1}) and (\ref{C2}), we find
\begin{eqnarray}
\Pi_{\mu} (E_1, E_2, q_{max}^2) &=& 2 N_c g_{\mu 0}\int 
{G_{E_1}^{(1)}({\bf x} = 0, {\bf p})G_{E_2}^{(2)}({\bf p}, {\bf y} = 0)}
\frac{d {\bf p}} {(2\pi )^3}. \nonumber 
\\
&=& 2 N_c g_{\mu 0}\int G_{E_1}^{(1)}(0, z)G_{E_2}^{(2)}(z, 0) d^3 z.
\end{eqnarray}
For the Green's function we use the representation
\begin{equation}
G_E ({\bf x, y}) = \sum_{l, m}\left (\sum_{n = l + 1}^{\infty}
\frac{\Psi_{nlm}({\bf x})\Psi_{nlm}^{*}({\bf y})}{E_{nl} - E - i0} +
\int 
\frac{d{ k}}{(2\pi)}\frac{\Psi_{klm}({\bf
x})\Psi_{klm}^{*}({\bf y})}
{{k} - E - i0}\right ).
\end{equation}
Provided ${\bf x} = 0$, only the terms with $l = 0$ are retained in the sum.
Then for the spectral density one has 
\begin{eqnarray}
\rho_{\mu}(E_1, E_2, q_{max}^2)=-2 N_c g_{\mu 0}\Psi_1^C (0)\Psi_2^C (0)
\int \tilde \Psi_{1E_1}^C ({\bf p}) \tilde \Psi_{2E_2}^C ({\bf p}) \cdot 
\frac{d {\bf p}} {(2\pi )^3},
\end{eqnarray}
where $\Psi_i^C$ are the Coulomb wave functions for the $(b\bar c)$ or $(c\bar
c)$ systems. An analogous expression can also be derived in the Born
approximation:
\begin{eqnarray}
\rho_{\mu}^B(E_1, E_2, q_{max}^2)=- 2 N_c g_{\mu 0}\Psi_1^f (0)\Psi_2^f (0)
\int \tilde \Psi_{1E_1}^f ({\bf p}) \tilde \Psi_{2E_2}^f ({\bf p}) \cdot 
\frac{d {\bf p}} {(2\pi )^3}.
\end{eqnarray}
Here $\Psi_i^f$ stands for the function of free quark motion. Since
the continuous spectrum Coulomb functions have the same normalization as
the free states, we obtain the approximation
\begin{equation}
\rho_{\mu}(E_1, E_2, q_{max}^2) \approx \rho_{\mu}^B(E_1, E_2, q_{max}^2)
\frac{\Psi_1^C (0)\Psi_2^C (0)}{\Psi_1^f (0)\Psi_2^f (0)}
\equiv\rho_{\mu}^B (E_1, E_2, q_{max}^2){\bf C},   
\label{r1}
\end{equation}
\begin{equation}
{\bf C} = \left \{\frac{4\pi\alpha_s}{3v_{13}}\left [1 - \exp\left (
-\frac{4\pi\alpha_s}{3v_{13}}\right )^{-1}\right ]
\frac{4\pi\alpha_s}{3v_{23}}\left [1 - \exp\left (
-\frac{4\pi\alpha_s}{3v_{23}}\right )^{-1}\right ]\right
\}^{\frac{1}{2}},\label{C}
\end{equation}
where $v_{13}$, $v_{23}$ are relative velocities in the $(b\bar c)$ and $(c\bar
c)$ systems, respectively. For them we have the following expressions:
\begin{eqnarray}
v_{13} &=& \sqrt{1 - \frac{4m_1m_3}{p_1^2 - (m_1 - m_3)^2}},\\
v_{23} &=& \sqrt{1 - \frac{4m_2m_3}{p_2^2 - (m_2 - m_3)^2}}.
\end{eqnarray}
Eq.(\ref{r1}) is exact for the identical quarkonia in the initial and final
states of transition under consideration. However, if the reduced masses are
different, then the overlapping of Coulomb functions can deviate from unity,
which breaks the exact validity of (\ref{r1}). From a pessimistic viewpoint,
this relation can serve as the estimate of upper bound on the form factor at
zero recoil. In reality, this boundary is practically saturated, which means
that in sum rules at low momenta inside the quarkonia, i.e. in the region of
physical resonances, the most essential effect comes from the normalization
factor $\bf C$, determined by the Coulomb function at the origin. The latter
renormalizes the coupling constant in the quark-mesonic vertex from the bare
value to the ``dressed" one. After that, the motion of heavy quarks in the
triangle loop is very close to that of free quarks.

In accordance with (\ref{R1}) for the Lorentz decomposition of $\rho_{\mu}
(p_1, p_2, q)$ we have 
\begin{equation}
\rho_{\mu}(p_1, p_2, q^2) = (p_1 + p_2)_{\mu}\rho_{+}(q^2) +
q_{\mu}\rho_{-}(q^2).
\end{equation}
As we have seen, the nonrelativistic expression of $\rho_\mu (E_1, E_2,
q_{max}^2)$ is proportional to the vector $(g_{\mu 0})$, which allows us to
isolate the evident combination of form factors $f_{\pm}$. The relations
between the form factors appearing in NRQCD at the recoil momentum close to
zero will be considered below. Here we stress only that we have
\begin{equation}
\rho_{+}(q_{max}^2) = \rho_{+}^B(q_{max}^2){\bf C},
\label{AR}
\end{equation} 
where the factor ${\bf C}$ has been specified in (\ref{C}). 

In the case of $B_c\to J/\psi l\nu_l$ transition, one can easily obtain
an analogous result for $\rho_0^A(q^2)$ (note that the form factor $F_0^A\sim
\rho_0^A$ gives the dominant contribution to the width of this decay
\cite{Jenkins}). In the nonrelativistic approximation we have
\begin{equation}
\rho_0^A(q_{max}^2) = \rho_0^{A,B} (q_{max}^2){\bf C}.\label{AC}
\end{equation}
To conclude this section, note that the derivation of formulas (\ref{AR}) and
(\ref{AC}) is purely formal, since the spectral densities are not specified
at $q_{max}^2$ (one can easily show $\rho_i^B (q^2)$ to be singular in this
point). Therefore, the resultant relations are valid only for $q^2$ approaching 
$q_{max}^2$. Unfortunately, this derivation does not give the $q^2$
dependence of the factor ${\bf C}$. But we suppose that ${\bf C}$ does not
crucially effect the pole behaviour of the form factors. Therefore, the
resultant widths of transitions can be treated as the saturated upper bounds in
the QCD sum rules.

\subsection{Gluon condensate contribution}

In this subsection we will discuss the calculation of Borel transformed
Wilson coefficient of the gluon condensate operator for the three-point sum
rules with arbitrary masses. The technique used
is the same as in \cite{Ball2} with some modifications  to simplify the 
resulting expression. As was noted in \cite{Ball2}, this method does not
allow for the subtraction of continuum contributions, which, however,
only a little change our results as the total contribution of the gluon
condensate to the three point sum rule is small by itself ($\leq 10\%$),
and, thus, its continuum portion is small, too. The form of the obtained
expression does not permit us to use the same argument as in \cite{Ball2} to
argue on the absence of those contributions at all. Their argument was based on
an expectation, that the typical continuum contribution can show up as
incomplete $\Gamma$ functions in the resulting expression and the absence of
them in the final answer leads authors of \cite{Ball2} to the conclusion, that
such contributions are actually absent in the processes, they considered.   

\begin{figure}[ph]
\begin{center}
\begin{picture}(150,180)
\put(0,120){\epsfxsize=3cm \epsfbox{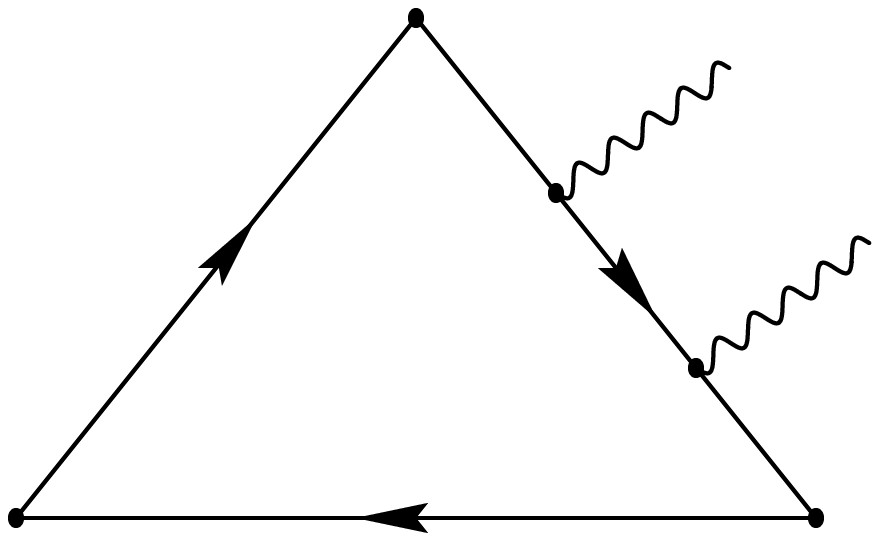}}
\put(75,120){\epsfxsize=3cm \epsfbox{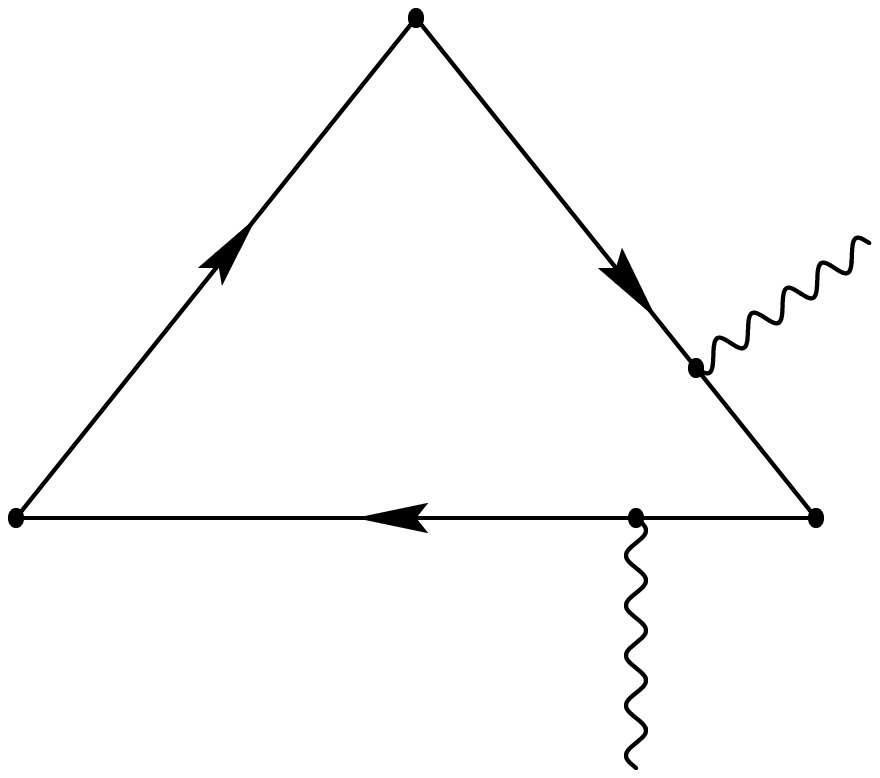}}
\put(0,60){\epsfxsize=3cm \epsfbox{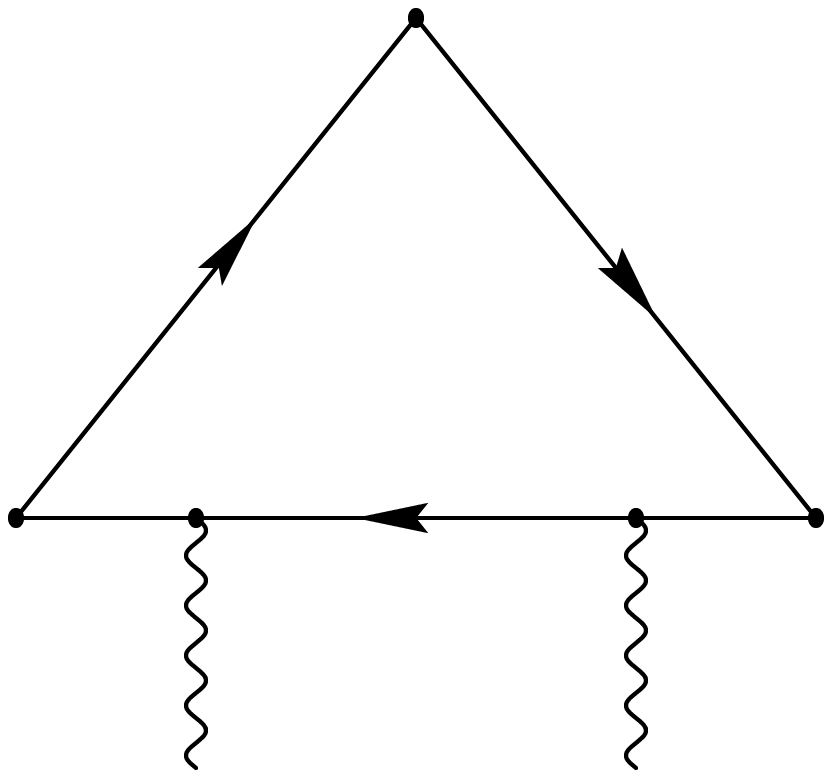}}
\put(75,60){\epsfxsize=3cm \epsfbox{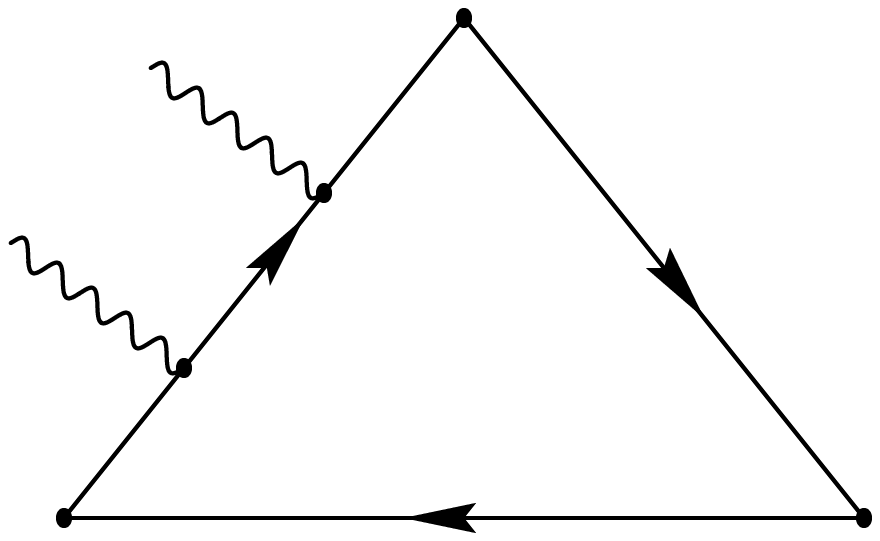}}
\put(0,0){\epsfxsize=3cm \epsfbox{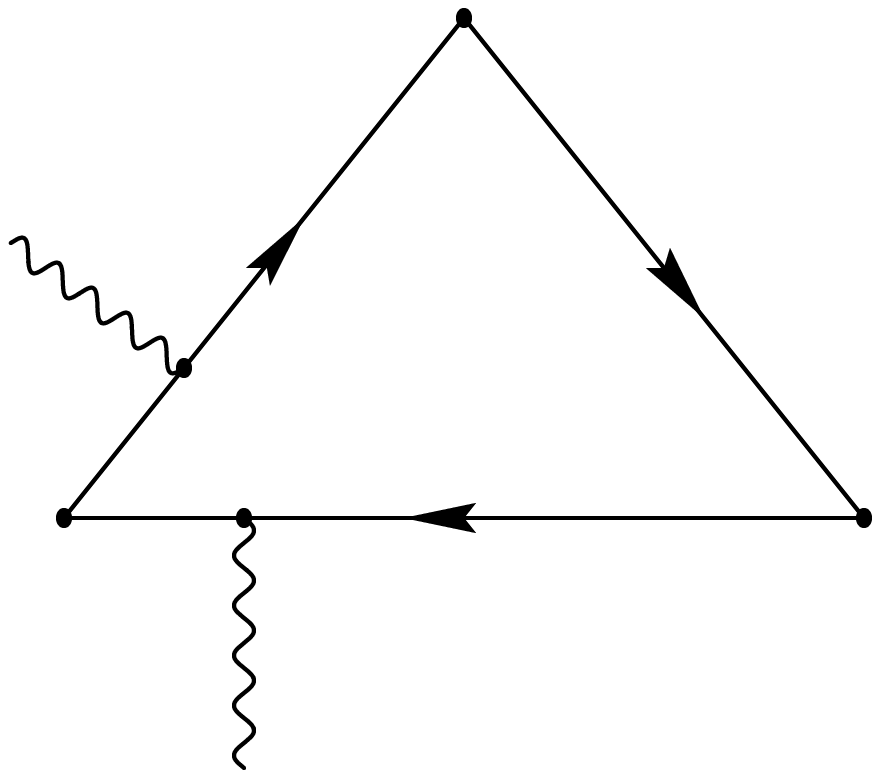}}
\put(75,0){\epsfxsize=3cm \epsfbox{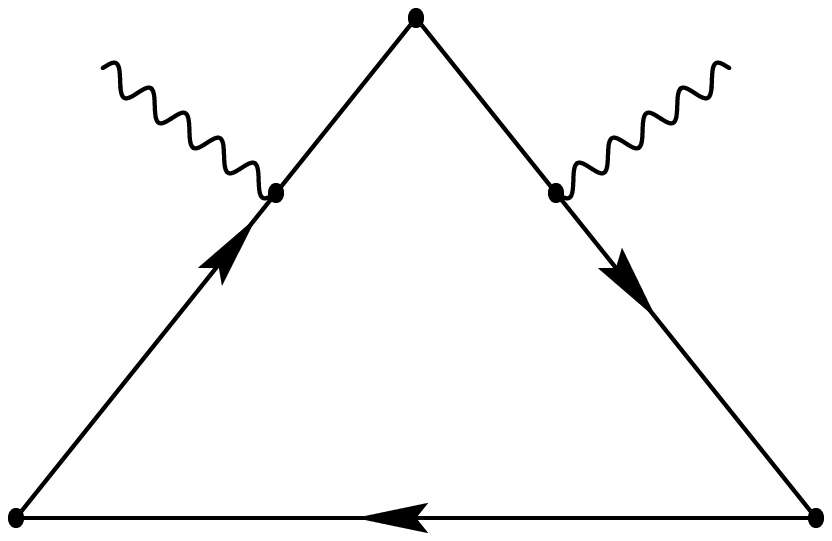}}

\put(0,140){$p_1$}
\put(60,140){$p_2$}
\put(30,135){$k$}
\put(30,176){$0$}
\put(62,160){$k_1$}
\put(52,172){$k_2$}
\put(6,158){$p_1+k$}
\put(54,148){$p_2+k$}

\put(0,80){$p_1$}
\put(60,80){$p_2$}
\put(30,75){$k$}
\put(30,116){$0$}
\put(16,62){$k_1$}
\put(43,62){$k_2$}
\put(-3,98){$p_1+k+k_1$}
\put(46,98){$p_2+k-k_2$}

\put(0,20){$p_1$}
\put(60,20){$p_2$}
\put(30,15){$k$}
\put(30,56){$0$}
\put(16,2){$k_1$}
\put(-2,42){$k_2$}
\put(46,38){$p_2+k$}

\put(75,20){$p_1$}
\put(135,20){$p_2$}
\put(105,15){$k$}
\put(105,56){$0$}
\put(82,52){$k_1$}
\put(126,52){$k_2$}
\put(78,38){$p_1+k$}
\put(121,38){$p_2+k$}

\put(75,80){$p_1$}
\put(135,80){$p_2$}
\put(105,75){$k$}
\put(105,116){$0$}
\put(82,112){$k_2$}
\put(73,102){$k_1$}
\put(121,98){$p_2+k$}

\put(75,140){$p_1$}
\put(135,140){$p_2$}
\put(105,135){$k$}
\put(105,176){$0$}
\put(137,160){$k_2$}
\put(118,122){$k_1$}
\put(81,158){$p_1+k$}

\end{picture}
\end{center}
\caption{The gluon condensate contribution to three-point QCD sum rules. The
directions of $p_1,\; k_1,\; k_2$ momenta are incoming, and that of $p_2$ is
outgoing.}
\label{cond}
\end{figure}

The gluon condensate contribution to three-point sum rules is given by
diagrams, depicted in Fig. \ref{cond}. For calculations we have used the
Fock-Schwinger
fixed point gauge \cite{Fock,Schw}:
\begin{equation}
x^{\mu}A_{\mu}^a (x) = 0,
\end{equation}
where $A_{\mu}^a$, $a = \{1, 2, ...,8\}$ is the gluon field.

In the evaluation of diagrams in Fig. \ref{cond} we encounter integrals of the
type\footnote{Since the diagrams under consideration do not have UV
divergencies, there is no need for a dimensional regularization.}
\begin{eqnarray}
I_{\mu_1\mu_2...\mu_n}(a,b,c) = \int\frac{d^4 k}{(2\pi)^4}
\frac{k_{\mu_1}k_{\mu_2}...k_{\mu_n}}{[k^2 - m_3^2]^a[(p_1 + k)^2 - m_1^2]^b
[(p_2 + k)^2 - m_2^2]^c}.
\end{eqnarray}
Continuing to Euclidean space-time and employing the Schwinger representation
for propagators, 
\begin{equation}
\frac{1}{[p^2 + m^2]^a} = \frac{1}{\Gamma (a)}\int_0^{\infty}
d\alpha\alpha^{a-1}e^{-\alpha (p^2 + m^2)}, 
\end{equation}
we find the following expression for the scalar integral ($n = 0$):
\begin{eqnarray}
I_0 (a, b, c) &=& \frac{(-1)^{a + b + c}i}{\Gamma (a)\Gamma (b)\Gamma (c)}
\int_0^{\infty}\int_0^{\infty}\int_0^{\infty} d\alpha d\beta d\gamma
\alpha^{a-1}\beta^{b-1}\gamma^{c-1}\nonumber\\
&& \int\frac{d^4 k}{(2\pi )^4}
e^{-\alpha (k^2 + m_3^2) - \beta (s_1 + k^2 + 2p_1\cdot k + m_1^2) - 
\gamma (s_2 + k^2 + 2p_2\cdot k + m_2^2)} .
\end{eqnarray}
This representation proves to be very convenient for applying the
Borel transformation with
\begin{equation}
\hat B_{p^2}(M^2)e^{-\alpha p^2} = \delta (1 - \alpha M^2).
\end{equation}
Then, we have
\begin{eqnarray}
\hat I_0 (a, b, c) &=& \frac{(-1)^{a + b + c}i}{\Gamma (a)\Gamma (b)\Gamma (c)
16\pi^2} (M_{cc}^2)^{2 - a -c}(M_{bc}^2)^{2 - a - b}\cdot\nonumber\\
&& U_0 (a + b + c - 4,1 - c -b),\\
\hat I_{\mu} (a, b, c) &=& \frac{(-1)^{a + b + c + 1}i}{\Gamma (a)\Gamma (b)
\Gamma (c)16\pi^2}\left (\frac{p_{1\mu}}{M_{bc}^2} + \frac{p_{2\mu}}{M_{cc}^2}
\right )(M_{cc}^2)^{3 - a - c}(M_{bc}^2)^{3 - a - b}\cdot\nonumber\\
&& U_0 (a + b + c - 5, 1 - c - b),\\ 
\hat I_{\mu\nu} (a, b, c) &=& \frac{(-1)^{a + b + c}i}{\Gamma (a)\Gamma (b)
\Gamma (c)16\pi^2}\left (\frac{p_{1\mu}}{M_{bc}^2} + \frac{p_{2\mu}}{M_{cc}^2}
\right )\cdot\nonumber \\
&& \left (\frac{p_{1\nu}}{M_{bc}^2} + \frac{p_{2\nu}}{M_{cc}^2}
\right )(M_{cc}^2)^{4 - a - c}(M_{bc}^2)^{4 - a - b}U_0 (a + b + c - 6,
1 - c - b) \nonumber \\
&& + \frac{g_{\mu\nu}}{2}\frac{(-1)^{a + b + c +1}i}{\Gamma (a)\Gamma (b)
\Gamma (c)16\pi^2}(M_{cc}^2)^{3 - a - c}(M_{bc}^2)^{3 - a - b}\cdot\\
&& U_0 (a + b + c - 6, 2 - c -b),\nonumber
\end{eqnarray}
where $M_{bc}^2$ and $M_{cc}^2$ are the Borel parameters in the $s_1$ and $s_2$
channels, respectively. Here we have introduced the $U_0(a,b)$ function,  which
is given by the following expression:
\begin{equation}
U_0 (a, b) = \int_0^{\infty} dy (y + M_{bc}^2 + M_{cc}^2)^ay^b
\exp \left [-\frac{B_{-1}}{y} - B_0 - B_1y\right ],
\end{equation}
where
\begin{eqnarray}
B_{-1} &=& \frac{1}{M_{cc}^2M_{bc}^2}(m_2^2M_{bc}^4 + m_1^2M_{cc}^4 +
M_{cc}^2M_{bc}^2(m_1^2 + m_2^2 - Q^2)), \nonumber\\
B_{0} &=& \frac{1}{M_{bc}^2M_{cc}^2}(M_{cc}^2(m_1^2 + m_3^2) + 
M_{bc}^2(m_2^2 + m_3^2))\label{CD},\\
B_{1} &=& \frac{m_3^2}{M_{bc}^2M_{cc}^2}. \nonumber
\end{eqnarray}
Then, one can express the results of calculation for any diagram in Fig.
\ref{cond} through $\hat I_0(a,b,c)$, $\hat I_{\mu}(a,b,c)$, $\hat
I_{\mu\nu}(a,b,c)$ and their derivatives over the Borel parameters, using the
partial fractioning of the integrand expression together with the following
relation:
\begin{eqnarray}
&& \hat B_{p_1^2}(M_{bc}^2)\hat B_{p_2^2}(M_{cc}^2)[p_1^2]^{m}[p_2^2]^{n}
I_{\mu_1\mu_2 ...\mu_n} (a,b,c) =\nonumber\\
&& [M_{bc}^2]^{m}[M_{cc}^2]^{n}\frac{d^m}{d(M_{bc}^2)^{m}}
\frac{d^n}{d(M_{cc}^2)^{n}}[M_{bc}^2]^{m}[M_{cc}^2]^{n}\hat 
I_{\mu_1\mu_2 ...\mu_n} (a,b,c).
\end{eqnarray}
The obtained expression can be further written down in terms of three
quantities
$\hat I_0(1,1,1)$, $\hat I_{\mu}(1,1,1)$,  $\hat I_{\mu\nu}(1,1,1)$ and 
their derivatives over the Borel parameters and quark masses by means of
\begin{equation}
\hat I_{\mu_1\mu_2 ...\mu_n}(a, b, c) = \frac{1}{\Gamma (a)
\Gamma (b)\Gamma (c)}\frac{d^{a - 1}}{d (m_3^2)^{a - 1}}
\frac{d^{b - 1}}{d (m_1^2)^{b-1}}\frac{d^{c - 1}}{d (m_2^2)^{c-1}}
\hat I_{\mu_1\mu_2 ...\mu_n}(1, 1, 1).
\end{equation} 

However, contrary to the case, discussed in \cite{Ball2}, in such calculations
the values of parameters $a, b$  will arise, for which the $U_0(a, b)$-function
has no analytical expression (it is connected to nonzero $m_3$ mass in our
case). The analytical approximations for $U_0(a, b)$ at these values of
parameters lead to very cumbersome expressions. The search for the most compact
form of the final answer leads to the conclusion, that the best decision in
this case is to express the result in terms of $U_0(a, b)$-function at
different values of its parameters. For this purpose we have used the following
transformation properties of $U_0(a, b)$:
\begin{eqnarray}
\frac{d U_0(a, b)}{d M_{bc}^2} &=& aU_0(a - 1, b) - 
\left (\frac{m_2^2}{M_{cc}^2} -  \frac{m_1^2M_{cc}^2}{M_{bc}^4}\right )
U_0(a, b - 1) + \nonumber\\
&& \frac{m_1^2 + m_3^2}{M_{bc}^4}U_0(a, b) + 
\frac{m_3^2}{M_{bc}^4M_{cc}^2}U_0(a, b + 1),\\
\frac{d U_0(a, b)}{d M_{cc}^2} &=& aU_0(a - 1, b) - 
\left (\frac{m_1^2}{M_{bc}^2} -  \frac{m_2^2M_{bc}^2}{M_{cc}^4}\right )
U_0(a, b - 1) + \nonumber\\
&& \frac{m_2^2 + m_3^2}{M_{cc}^4}U_0(a, b) + 
\frac{m_3^2}{M_{bc}^2M_{cc}^4}U_0(a, b + 1).
\end{eqnarray}  
In Appendix B we have presented an analytical expression, obtained 
in this way, for the Wilson coefficient of gluon condensate operator, 
contributing to the $\Pi_1 = \Pi_{+} + \Pi_{-}$ amplitude. One can
see that, even in this form the obtained results are very cumbersome.
So, we have realized the gluon condensate corrections as C++ codes, where
the functions $U_0(a, b)$ are evaluated numerically. Analytical approximations,
which can be made for the $U_0(a, b)$ functions are discussed in Appendix B.

In Fig. \ref{qcdf+} we have shown the effect of gluon condensate on the
$f_{1}(0)$ form factor in the Borel transformed three-point sum rules. 

\setlength{\unitlength}{1mm}

\begin{figure}[th]

\vspace*{-2.5cm}
\begin{center}
\begin{picture}(140,100)
\put(0,10){
\epsfxsize=10cm
\epsfbox{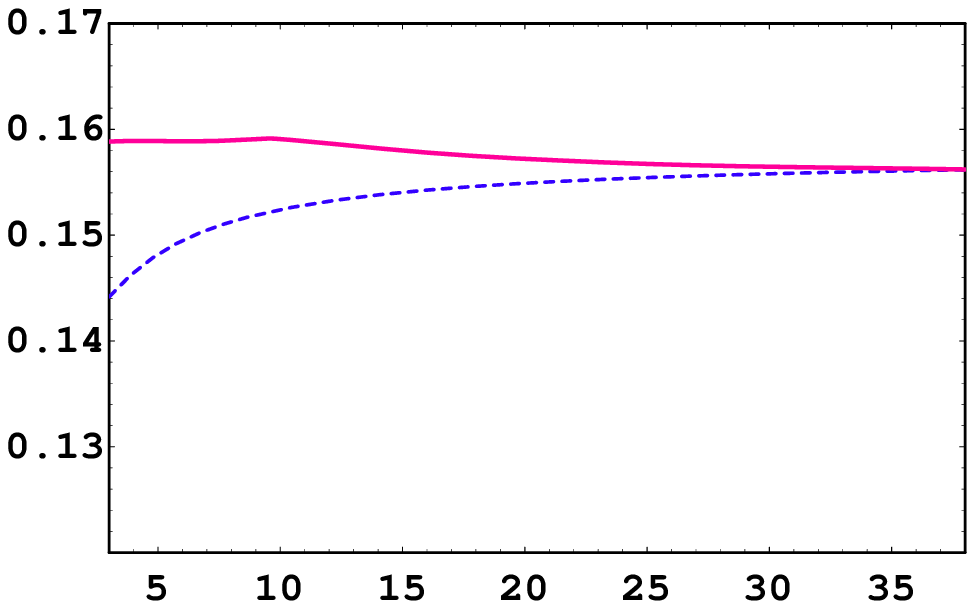}}
\put(100,10){$M^2_{cc}$, GeV$^2$}
\put(10,75){$f_1(0)$}
\end{picture}
\end{center}

\vspace*{-1cm}
\caption{The gluon condensate contribution to the $f_{1}(0)$ form factor in
the Borel-transformed sum rules at fixed $M^2_{bc}=70$ GeV$^2$. The dashed line
represent the bare quark-loop results, and the solid curve is the form factor
including the gluon condensate term at $\langle \frac{\alpha_s}{\pi} G^2\rangle
=10^{-2}$ GeV$^4$.}
\label{qcdf+}
\end{figure}
\normalsize

We can draw the conclusion that the calculation of gluon condensate term in
full QCD sum rules allows one to enlarge the stability region in the parameter
space for the form factors, which indicates the reliability of sum rules
technique.

\subsection{Numerical results on the form factors}

First, we evaluate the form factors in the scheme of spectral density moments.
This scheme is not strongly sensitive to the values of threshold energies,
determining the region of resonance contribution. In the calculations we put
\begin{eqnarray}
k_{th}(\bar b c) & = & 1.5\; {\rm GeV},\nonumber \\
k_{th}(\bar c c) & = & 1.2\; {\rm GeV}, \\
m_b & = & 4.6\; {\rm GeV},\nonumber \\
m_c & = & 1.4\; {\rm GeV},\nonumber 
\end{eqnarray}
where $k_{th}$ is the momentum of quark motion in the rest frame of quarkonium.
The chosen values of threshold momenta correspond to the minimal energy of
heavy meson pairs in specified channels. 

The typical behaviour of form factors in the moment scheme of QCD sum rules is
presented in Fig. \ref{f2mom}.

The evaluation of Coulomb corrections strongly depends on the appropriate set
of $\alpha_s$ for the quarkonia under consideration. The corresponding scale of
gluon virtuality is determined by quite a low value close to the average
momentum transfer in the system. So, the expected $\alpha_s$ is about 0.5. To
decrease the uncertainty we consider the contribution of Coulomb rescattering
in the two-point sum rules giving the leptonic constants of heavy quarkonia. 
These sum rules are quite sensitive to the value of strong coupling constant,
as the perturbative contribution depends on it linearly.
The observed value for the charmonium, $f_\psi\approx 410$ MeV, can be obtained
in this technique at $\alpha_s^{coul}(\bar c c) = 0.6$. The value 
$f_{B_c}=385$ MeV, 
as it is predicted in QCD sum rules \cite{fbc}, gives $\alpha_s^{coul}(\bar b 
c) = 0.45$. We present the results of the 
Coulomb enhancement for the form factors 
in Table \ref{form}.
\setlength{\unitlength}{1mm}

\begin{figure}[ph]

\vspace*{-2cm}
\begin{center}
\begin{picture}(140,100)
\put(0,-120){
\epsfxsize=10cm
\epsfbox{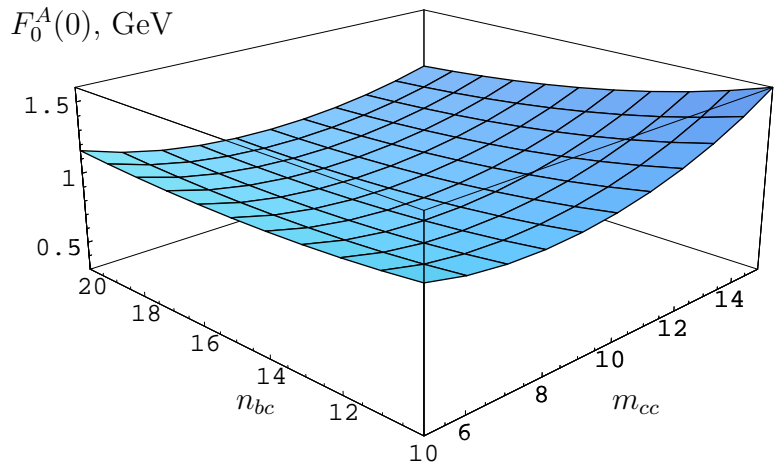}}
\put(0,60){$F_{0}^A(0)$, GeV}
\put(30,10){$n_{bc}$}
\put(80,10){$m_{cc}$}
\end{picture}
\end{center}
\caption{The QCD sum rule results in the moment scheme for the $F_{0}^A(0)$
form factor in the bare approximation.}
\label{f2mom}
\end{figure}
\normalsize

\begin{figure}[ph]

\vspace*{-2cm}
\begin{center}
\begin{picture}(140,100)
\put(0,-120){
\epsfxsize=10cm
\epsfbox{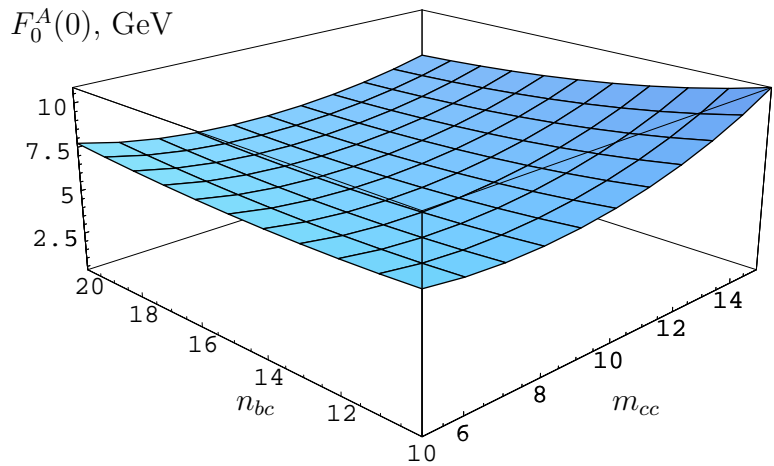}}
\put(0,60){$F_{0}^A(0)$, GeV}
\put(30,10){$n_{bc}$}
\put(80,10){$m_{cc}$}
\end{picture}
\end{center}
\caption{The QCD sum rule results in the moment scheme for the $F_{0}^A(0)$
form factor with account of Coulomb corrections.}
\label{f3mom}
\end{figure}
\normalsize

\noindent
The result after the introduction of the 
Coulomb correction is shown in Fig. \ref{f3mom}.
Such large corrections to the form factors should not lead to a 
confusion, as they are resulted from the fare account of the Coulomb 
corrections both for bare quark loop diagram and meson coupling constant.  

\begin{figure}[th]

\vspace*{-2cm}
\begin{center}
\begin{picture}(140,100)
\put(4,-160){
\epsfxsize=10cm
\epsfbox{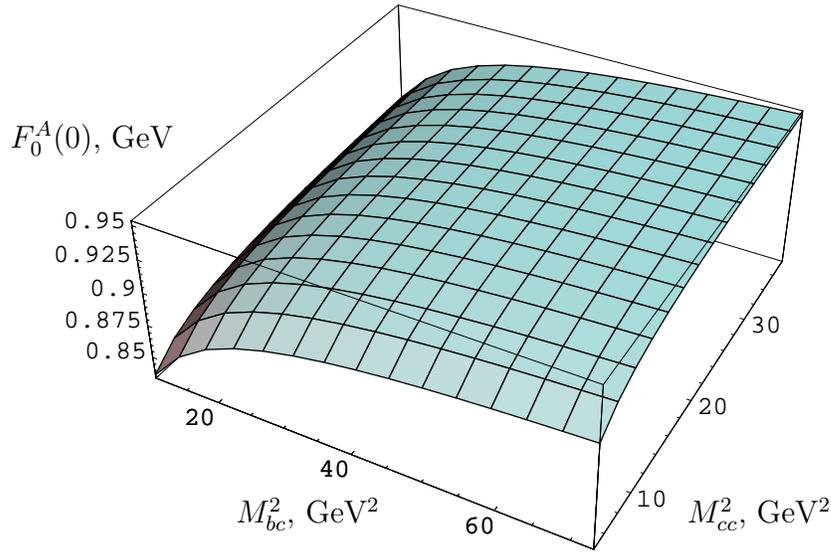}}
\put(0,60){$F_{0}^A(0)$, GeV}
\put(30,10){$M^2_{bc}$, GeV$^2$}
\put(90,10){$M^2_{cc}$, GeV$^2$}
\end{picture}
\end{center}
\caption{The Borel transformed sum rule results for the $F_{0}^A(0)$ form
factor in the bare approximation of QCD.}
\label{f4bor}
\end{figure}

\begin{figure}[th]

\vspace*{-2cm}
\begin{center}
\begin{picture}(140,100)
\put(7,-160){
\epsfxsize=10cm
\epsfbox{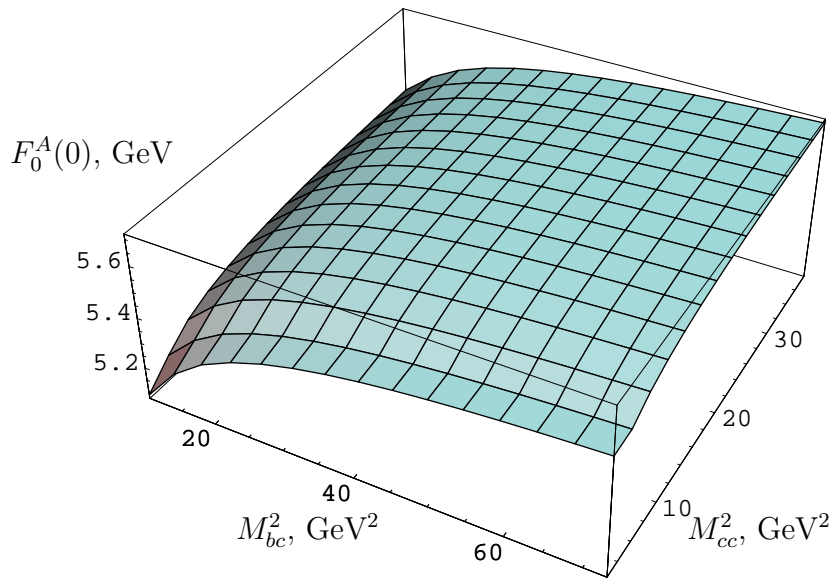}}
\put(0,60){$F_{0}^A(0)$, GeV}
\put(30,10){$M^2_{bc}$, GeV$^2$}
\put(90,10){$M^2_{cc}$, GeV$^2$}
\end{picture}
\end{center}
\caption{The Borel transformed sum rule results for the $F_{0}^A(0)$ form
factor with account of Coulomb corrections.}
\label{f5bor}
\end{figure}

In the scheme of the Borel transformation we find a strong dependence on the
thresholds of continuum contribution. We think that this dependence reflects
the influence of contributions coming from the excited states. So, the choice
of $k_{th}$ values in the same region as in the scheme of spectral density
moments results in the form factors, which are approximately 50\% greater than
the predictions in the moments scheme, where the higher excitations numerically
are not essential. In this case we can explore the ideology of finite energy
sum rules \cite{fesr}, wherein the choice of interval for the quark-hadron
duality, expressed by means of sum rules, allows one to isolate the
contribution of basic states only. So, if we put
\begin{eqnarray}
k_{th}(\bar b c) & = & 1.2\; {\rm GeV},\nonumber \\
k_{th}(\bar c c) & = & 0.9\; {\rm GeV}, 
\end{eqnarray}
then the region of the lowest bound states is taken into account in both
channels of initial and final states, and the Borel transform scheme leads to
the results, which are very close to those of moment scheme. The dependence of
calculated values on the Borel parameters is presented in Figs. \ref{f4bor} and
\ref{f5bor}, in the bare and Coulomb approximations, respectively.
\begin{table}[th] 
\begin{center} 
\begin{tabular}{|l|l|l|l|l|l|l|} 
\hline
approx. & $f_+$ & $f_-$ & $F_V$, GeV$^{-1}$ & $F_0^A$, GeV  
& $F_+^A$, GeV$^{-1}$ & $F_-^A$, GeV$^{-1}$ \\ 
\hline
bare  & 0.10 & -0.057 & 0.016 & 0.90 & -0.011 & 0.018 \\ 
\hline 
coul. & 0.66 & -0.36  & 0.11  & 5.9  & -0.074 & 0.12  \\
\hline 
\end{tabular} 
\end{center} 
\caption{The form factors of $B_c^+$ decay modes into the heavy quarkonia 
at $q^2=0$ in the bare quark-loop approximation taking into 
 account for the Coulomb  correction.} 
\label{form} 
\end{table} 

\normalsize

As for the dependence of form factors on $q^2$, the consideration of bare quark
loop term shows that, say, for $F_0^A(q^2)$ it can be approximated by the pole 
function: 
\begin{equation} 
F_0^A(q^2) = \frac{F_0^A(0)}{1-\frac{q^2}{M^2_{pole}}}, 
\end{equation} 
with $M_{pole} \approx 4.5$ GeV. The latter is in a good agreement with the 
value given in \cite{bagan}. However, we believe that the pole mass can change 
after the inclusion of $\alpha_s$ corrections\footnote{In HQET, the slope of 
Isgur-Wise function acquires a valuable correction due to the 
$\alpha_s$-term.}. From the naive meson dominance model we expect that
$M_{pole} \approx 6.3 - 6.5$ GeV. 
\normalsize

We have calculated the total widths of semileptonic decays in the region of 
$M_{pole}= 4.5 - 6.5$ GeV, which result in the 30\% variation of predictions 
for the modes with the massless leptons and more sizable dependence for the 
modes with the $\tau$ lepton (see Table \ref{tot}). 

To compare with other estimates we calculate the width of $\bar b\to \bar c
e^+\nu$ transition as the sum of decays into the pseudoscalar and vector states
and find\footnote{For normalization of the calculated  branching ratios 
we used the total $B_c$ width obtained in the framework of the OPE approach
[8].} 
BR$(B_c^+\to c\bar c e^+\nu)\approx 3.4\pm 0.6$ \%, which is in a good
agreement with the value obtained in potential models \cite{pm} and in OPE
calculations \cite{beneke}, where the following estimate was obtained
BR$(B_c^+\to c\bar c e^+\nu)\approx 3.8$ \%.

In the presented results we have supposed the quark mixing matrix element
$|V_{bc}|=0.040$.

\begin{table}[th] 
\begin{center} 
\begin{tabular}{|l|l|l|} 
\hline 
mode & $\Gamma$, $10^{-15}$ GeV & BR, \% \\ 
\hline 
$\eta_c e^+ \nu_e$ & $11\pm 1$ & $0.9\pm 0.1$\\ 
\hline 
$\eta_c \tau^+ \nu_\tau$ & $3.3\pm 0.9$ & $0.27\pm 0.07$\\ 
\hline 
$\psi e^+ \nu_e$ & $28\pm 5$ & $2.5\pm 0.5$\\
\hline 
$\psi \tau^+ \nu_\tau$ & $7\pm 2$ & $0.60\pm 0.15$\\
\hline 
\end{tabular} 
\end{center}
\caption{The width of $B_c^+$ decay modes into the heavy quarkonia and leptonic 
pair and the branching fractions, calculated at $\tau_{B_c}=0.55$ ps.} 
\label{tot} 
\end{table} 

As for the hadronic decays, in the approach of factorization \cite{blokshif} we
assume that the width of transition $B_c^+\to J/\psi (\eta_c)+$ {\it light
hadrons} can be calculated with the same form factors after the introduction of
QCD corrections, which can be easily written down as the factor $H= N_c
a_1^2$. The factor $a_1$ represent the hard $\alpha_s$ corrections to the four
fermion weak interaction. Numerically we put $a_1=1.2$, which yields $H \approx 
4.3$. So, we find
\begin{eqnarray}
{\rm BR}[B_c^+\to J/\psi+\mbox{\it light hadrons}] & = & 11\pm 2 \%,\\
{\rm BR}[B_c^+\to \eta_c+\mbox{\it light hadrons}] & = & 4.0\pm 0.5 \%.
\end{eqnarray}
Neglecting the decays of $\bar b\to \bar c c \bar s$, which are suppressed by
both the small phase space and the negative Pauli interference of decay product
with the charmed quark in the initial state \cite{beneke}, we evaluate the
branching fraction of beauty decays in the total width of $B_c$ as
$$
{\rm BR}[B_c^+\to \bar c c+\mbox{\it X}]  =  23\pm 5 \% ,
$$
which is in agreement with the estimates in other approaches \cite{pm,beneke},
where this value is equal to 25 \%.

\section{Three-point NRQCD sum rules}

The formulation of sum rules in NRQCD follows the same lines as in QCD, the
only difference is the lagrangian, describing strong interactions of heavy
quarks.
\subsection{Symmetry of form factors in NRQCD and one-loop approximation}

At the recoil momentum close to zero, the heavy quarks in both the initial and
final states have small relative velocities, so that the dynamics of heavy
quarks is essentially nonrelativistic. This allows us to use the NRQCD
approximation in the study of mesonic form factors. As in the case
of heavy quark effective theory (HQET), the expansion in the small relative
velocities to the leading order leads to various relations between the
different form factors. Solving these relations results in the introduction of
an universal form factor (an analogue of the Isgur-Wise function) at $q^2\to
q^2_{max}$.

In this subsection we consider the limit
\begin{eqnarray}
v_1^{\mu} &\neq & v_2^{\mu}, \label{cs}\\ 
w &=& v_1\cdot v_2\to 1,\nonumber
\end{eqnarray}
where $v_{1,2}^{\mu} = p_{1,2}^{\mu}/\sqrt{p_{1,2}^2}$ are the four-velocities
of heavy quarkonia in the initial and final states. The study of region
(\ref{cs}) is reasonable enough, because in the rest frame of $B_c$-meson 
($p_1^{\mu} = (\sqrt{p_1^2}, \vec 0)$), the four-velocities differ only by a
small value $|\vec p_2|$ $(p_2^{\mu} = (\sqrt{p_2^2}, \vec p_2)$, whereas their
scalar product $w$ deviates from unity only due to a term, proportional to the
square of $|\vec p_2|$: 
$w = \sqrt{1 + \frac{|\vec p_2|^2}{p_2^2}}\sim 1 + \frac{1}{2}\frac{|\vec
p_2|^2}{p_2^2}$.
Thus, in the linear approximation at $|\vec p_2|\to 0$, relations
(\ref{cs}) are valid and take place.

Here we would like to note, that (\ref{cs}) generalizes the investigation of
\cite{Jenkins}, where the case of $v_1 = v_2$ was considered. This condition
severely restricts the relations of spin symmetry for the form factors and, as
a consequence, it provides a single connection between the form factors.

As can be seen in Fig. 1, since the antiquark line with the mass $m_3$ is 
common to the  heavy quarkonia, the four-velocity of antiquark can be
written down as a linear combination of four-velocities $v_1$ and $v_2$:
\begin{equation}
\tilde v_3^{\mu} = av_1^{\mu} + bv_2^{\mu}. \label{v3}
\end{equation}
In the leading order of NRQCD for the kinematical invariants, determining the
spin structure of quark propagators in the limit $w\to 1$, we have the
following expressions:
\begin{eqnarray}
p_1^2 &\to & (m_1 + m_3)^2 = {\cal M}_{1}^2,\nonumber \\
p_2^2 &\to & (m_2 + m_3)^2 = {\cal M}_{2}^2,\nonumber \\
\Delta_1 &\to & 2m_3{\cal M}_{1},\nonumber \\
\Delta_2 &\to & 2m_3{\cal M}_{2}, \\
u &\to & 2{\cal M}_{1}{\cal M}_{2}\cdot w,\nonumber \\
\lambda (p_1^2, p_2^2, q^2) &\to & 4{\cal M}_{1}^2{\cal M}_{2}^2 (w^2 -
1).\nonumber
\end{eqnarray}
In this kinematics it is an easy task to show, that in (\ref{v3}) $a = b =
-\frac{1}{2}$, i.e.
\begin{equation}
\tilde v_3^{\mu} = -\frac{1}{2}(v_1 + v_2)^{\mu}.
\end{equation}
Applying the momentum conservation in the vertices on Fig. 1 we derive the
following formulae for the four-velocities of quarks with the masses $m_1$ and
$m_2$:
\begin{eqnarray}
\tilde v_1^{\mu} &=& v_1^{\mu} + \frac{m_3}{2m_1}(v_1 - v_2)^{\mu}, \\
\tilde v_2^{\mu} &=& v_2^{\mu} + \frac{m_3}{2m_2}(v_2 - v_1)^{\mu},
\end{eqnarray}
and in the limit $w\to 1$, we have $\tilde v_1^2 = \tilde v_2^2 = 1$, as it
should be.

After these definitions have been done, it is straightforward to write down the
transition form factor for the current $J_{\mu} = \bar Q_1\Gamma_{\mu} Q_2$
with the spin structure $\Gamma_{\mu} = \{\gamma_{\mu}, \gamma_5\gamma_{\mu}\}$
\begin{eqnarray}
\langle H_{Q_1\bar Q_3}|J_{\mu}|H_{Q_2\bar Q_3}\rangle &=& 
tr[\Gamma_{\mu}\frac{1}{2}(1 + \tilde v_1^{\mu}\gamma_{\mu})\Gamma_1\frac{1}{2}
(1 + \tilde v_3^{\nu}\gamma_{\nu})\cdot \nonumber \\
&& \Gamma_2\frac{1}{2}(1 + \tilde v_2^{\lambda}\gamma_{\lambda})]\cdot 
h(m_1, m_2, m_3),
\end{eqnarray}
where $\Gamma_{1}$ determines the spin state in the heavy quarkonium $Q_1\bar
Q_3$ (in our case it is pseudoscalar, so that $\Gamma_{1} = \gamma_5$),
$\Gamma_2$ determines the spin wave function of quarkonium in the final
state: $\Gamma_2 = \{\gamma_5,  \epsilon^{\mu}\gamma_{\mu}\}$ for the
pseudoscalar and vector states, respectively  ($H = {P,V}$).  The quantity $h$
is an universal function at $w\to 1$, independent of the quarkonium spin state.
So, for the form factors, discussed in our paper, we
have
\begin{eqnarray}
\langle P_{Q_1\bar Q_3}|\bar Q_1\gamma^{\mu} Q_3|P_{Q_2\bar Q_3}\rangle &=& 
(c_1^{P}\cdot v_1^{\mu} + c_2^{P}\cdot v_2^{\mu})\cdot h, \\
\langle P_{Q_1\bar Q_3}|\bar Q_1\gamma^{\mu} Q_3|V_{Q_2\bar Q_3}\rangle &=& 
i c_V\cdot\epsilon^{\mu\nu\alpha\beta}\epsilon_{\nu}v_{1\alpha}v_{2\beta}\cdot
h,
\\
\langle P_{Q_1\bar Q_3}|\bar Q_1\gamma_5\gamma^{\mu} Q_3|V_{Q_2\bar Q_3}\rangle
&=& 
(c_{\epsilon}\cdot\epsilon^{\mu} + c_1\cdot v_1^{\mu}(\epsilon\cdot v_1) + 
c_2\cdot v_2^{\mu}(\epsilon\cdot v_1))\cdot h,
\end{eqnarray}
where
\begin{eqnarray}
c_{\epsilon} &=& -2,\nonumber\\
c_1 &=& -\frac{m_3(3m_1 + m_3)}{4m_1m_2},\nonumber\\
c_2 &=& \frac{1}{4m_1m_2}(4m_1m_2 + m_1m_3 + 2m_2m_3 + m_3^2),\nonumber\\
c_V &=& -\frac{1}{2m_1m_2}(2m_1m_2 + m_1m_3 + m_2m_3),\\
c_1^{P} &=& 1 + \frac{m_3}{2m_1} - \frac{m_3}{2m_2},\nonumber\\ 
c_2^{P} &=& 1 - \frac{m_3}{2m_1} + \frac{m_3}{2m_2}.\nonumber 
\end{eqnarray}
Then for the form factors in NRQCD we have the following symmetry relations:
\begin{eqnarray}
f_{+}(c_1^{P}\cdot{\cal M}_2 - c_2^{P}{\cal M}_1) - f_{-}(c_1^{P}\cdot{\cal
M}_2 + c_2^{P}\cdot{\cal M}_1) &=& 0,\nonumber\\
F_{0}^{A}\cdot c_V - c_{\epsilon}\cdot F_V{\cal M}_1{\cal M}_2 &=& 0,
\label{Fsym}\\
F_{0}^{A}(c_1 + c_2) - c_{\epsilon}{\cal M}_1 (F_{+}^{A}({\cal M}_1 + {\cal
M}_2) + 
F_{-}^{A}({\cal M}_1 - {\cal M}_2)) &=& 0,\nonumber \\
F_{0}^{A}c_1^{P} + c_{\epsilon}\cdot{\cal M}_1(f_{+} + f_{-}) &=& 0. \nonumber
\end{eqnarray}
Thus, we can claim, that in the approximation of NRQCD, the form factors of
weak currents responsible for the transitions between two heavy quarkonium
states are given in terms of the single form factor, say, $F_{0}^{A}$. The
exception is observed for the form factors $F_{+}^{A}$ and $F_{-}^{A}$, since
the definite value is only taken by their linear combination $F_{+}^{A}({\cal
M}_1 + {\cal M}_2) + F_{-}^{A}({\cal M}_1 - {\cal M}_2)$.
This fact has a simple physical explanation. Indeed, the polarization of
vector quarkonium $\epsilon^{\mu}$ has two components: the longitudinal term
$\epsilon_L^{\mu}$ and transverse one $\epsilon_T^{\mu}$ (i.e.
$(\epsilon_T\cdot v_1) = 0$). $\epsilon_L^{\mu}$ can be decomposed in terms of
$v_1^{\mu}$ and $v_2^{\mu}$:
\begin{equation}
\epsilon^{\mu} = \alpha\epsilon_L^{\mu} + \beta\epsilon_T^{\mu},\quad 
\alpha^2 + \beta^2 = 1, 
\end{equation}
where
\begin{eqnarray}
\epsilon_L^{\mu} &=& \frac{1}{\sqrt{s_2k}}(-2s_2p_1^{\mu} + up_2^{\mu})\to 
\frac{1}{\sqrt{w^2 - 1}}(-v_1^{\mu} + wv_2^{\mu}),\nonumber \\
\alpha &=& -\frac{2\sqrt{s_2}}{\sqrt{k}}(\epsilon_L\cdot p_1)\to 
-\frac{1}{\sqrt{w^2 - 1}}(\epsilon\cdot v_1). \label{xe}
\end{eqnarray}
From (\ref{xe}) one can see that the decomposition of polarization vector
$\epsilon$ into the longitudinal and transverse parts in NRQCD is singular in
the limit $w\to 1$:
\begin{equation}
\epsilon^{\mu} = -\frac{1}{w^2 - 1}(-v_1^{\mu} + wv_2^{\mu})(\epsilon\cdot v_1)
+
\beta\cdot\epsilon_T^{\mu}.
\end{equation}
It means, that the introduction to the form factor $F_{0}^{A}$ of an additional
term $\Delta F_0^{A} = (w^2 - 1)\cdot\delta h$, which vanishes at $w\to 1$ and,
thus, is not under control in NRQCD, leads to a finite correction for the
form factors $F_{+}^{A}$ and $F_{-}^{A}$. This correction is cancelled in the
special linear combination of form factors, presented in (\ref{Fsym}).

In the case of $v_1 = v_2$ we reproduce the single relation between form
factors $F_{0}^{A}$ and $f_{\pm}$, as it was obtained early in \cite{Jenkins}.

Thus, we have obtained the generalized relations due to the spin symmetry of
NRQCD lagrangian for the case $v_1\neq v_2$ in the limit, where the invariant
mass of lepton pair takes its maximum value, i.e at the recoil momentum close
to zero.

In the one-loop approximation for the three-point NRQCD sum rules, i.e. in the
calculation of bare quark loop, the symmetry relations (\ref{Fsym}) take place
already for the double spectral densities $\rho_j^{NR}$ in the limit $|\vec
p_2|\to 0$. We have checked, that the spectral densities of the full QCD in the
NRQCD limit $w\to 1$ satisfy the symmetry relations (\ref{Fsym}).

It is easily seen that in this approximation,
\begin{equation}
\rho_{0}^{A, NR} = -\frac{6m_1m_2m_3}{|\vec p_2|(m_1 + m_3)}.
\end{equation}
When integrating over the resonance region, we must take into account that
\begin{equation}
\frac{|\omega_2 - \omega_1\frac{m_{13}}{m_{23}}|m_2}{|\vec
p_2|\sqrt{2\omega_1m_{13}}} 
\leq 1,
\end{equation}
and  so we see, that in the limit $|\vec p_2|\to 0$ the integration region
tends to a single point. Here $p_1 = (m_1 + m_3 + \omega_1, \vec 0)$, $p_2 =
(m_2 +
m_3 + \omega_2, \vec
p_2)$ and
$m_{ij} = \frac{m_im_j}{m_i + m_j}$ is the reduced mass of system $(Q_i\bar
Q_j)$. 

\begin{figure}[ph]

\vspace*{-3cm}
\begin{center}
\begin{picture}(140,100)
\put(20,0){
\epsfxsize=5cm
\epsfbox{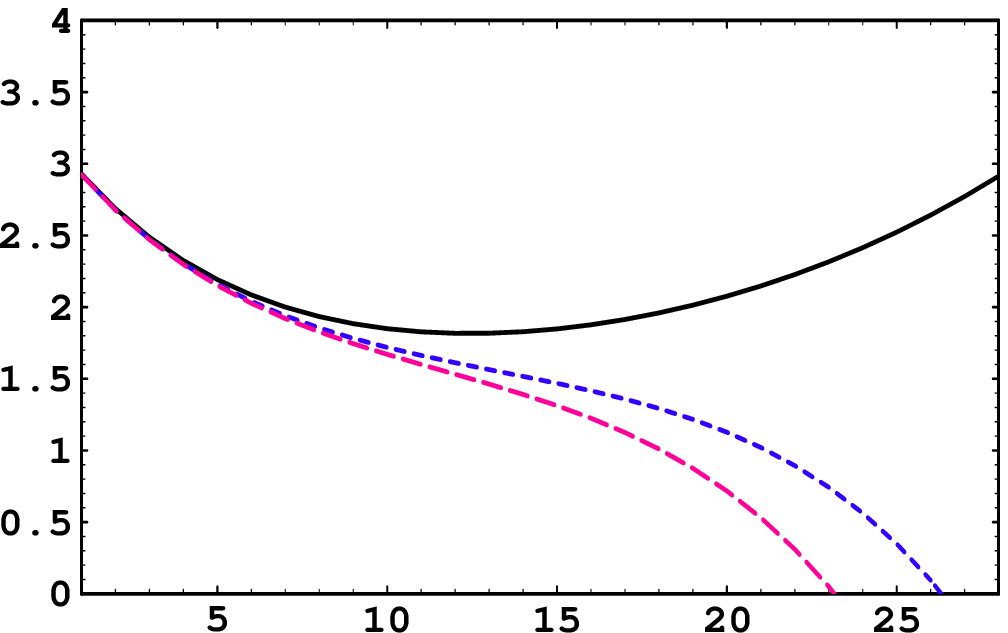}}
\put(100,-3){$m$}
\put(10,67){$F_0^A(0)$, GeV}
\end{picture}
\end{center}
\caption{The NRQCD sum rule results for the $F_{0}^A(0)$ form factor. The solid
line represents the bare quark loop contribution, the short dashed line is the
result obtained by taking into 
account  the gluon condensate term $R_0$ only, and the long
dashed line is the form factor including the full expression for the gluon
condensate.}
\label{nrfig1}
\end{figure}

After the substitutions of variables $\omega_1 = \frac{k^2}{2m_{13}}$ and 
$x = (\omega_2 - \frac{k^2}{2m_{23}})\cdot\frac{m_2}{|\vec p_2|k}$, in the
limit 
$|\vec p_2|\to 0$ for the correlator $\Pi_{0}^{A, NR}$ we have the following
expression:
\begin{eqnarray}
\Pi_{0}^{A, NR} &=& -\frac{1}{(2\pi)^2}\int\frac{d\tilde \omega_1 d\tilde
\omega_2}
{(\tilde \omega_1 - \omega_1)(\tilde \omega_2 - \omega_2)}\rho_{0}^{A, NR} =
\nonumber \\
&=& \frac{3}{\pi^2}\int_0^{k_{th}}\frac{k^2dk}{(\omega_1 - \frac{k^2}{2m_{13}})
(\omega_2 - \frac{k^2}{2m_{23}})},
\end{eqnarray}
where $k_{th}$ denotes the resonance region boundary. In the method of moments
in NRQCD sum rules we put $\omega_1 = -(m_1 + m_3) + q_1$ and $\omega_2 = -(m_2
+ m_3) + q_2$, so that in the limit $q_{1,2}\to 0$ we have
\begin{eqnarray}
\frac{1}{n!}\frac{1}{m!}\frac{d^{n + m}}{dq_1^ndq_2^m}\Pi_{0}^{A, NR} &=& 
\Pi_{0}^{A, NR}[n, m] \nonumber \\
&=& \frac{3}{\pi^2}\int_0^{k_{th}}\frac{k^2dk}{({\cal M}_{1} +
\frac{k^2}{2m_{13}})^{n+1}
({\cal M}_{2} + \frac{k^2}{2m_{23}})^{m+1}}.
\end{eqnarray}
In the hadronic part of NRQCD sum rules in the limit $|\vec p_2|\to 0$ we model
the resonance contribution by the following presentation:
\begin{equation}
\Pi_{0}^{A, res} = \sum_{i,j}\frac{f_i^{Q_1\bar Q_3}M_{1, i}^2}{(m_1 +
m_3)M_{1, i}^2}
\frac{f_j^{Q_2\bar Q_3}M_{2, j}}{M_{2, j}^2}F_{0,ij}^{A}\sum_{l,m}\left 
(\frac{q_1^2}{M_{1,i}^2}\right )^l\left (\frac{q_2^2}{M_{2,j}^2}\right )^m ,
\label{nrhad}
\end{equation}
Saturating the $p_1^2$ and $p_2^2$ channels by ground states of mesons under
consideration
we have 
\begin{equation}
F_{0, 1S\to 1S}^{A} = \frac{\Pi_{0}^{A, NR}[n, m](m_1 + m_3)}{f_{1S}^{Q_1\bar
Q_3}
M_{1,1S}^2 f_{1S}^{Q_2\bar Q_3}M_{2,1S}^2}M_{1, 1S}^nM_{2, 1S}^m.
\label{nrf0}
\end{equation}
The value of $F_{0, 1S\to 1S}^{A}$ at fixed $n=4$  is presented in Fig.
\ref{nrfig1} at
\begin{eqnarray}
k_{th} &=& 1.3\; {\rm GeV},\nonumber \\
m_b &=& 4.6\; {\rm GeV},\nonumber \\
m_c &=& 1.4\; {\rm GeV}.\nonumber
\end{eqnarray}

\begin{figure}[ph]

\vspace*{-3cm}
\begin{center}
\begin{picture}(140,100)
\put(20,0){
\epsfxsize=5cm
\epsfbox{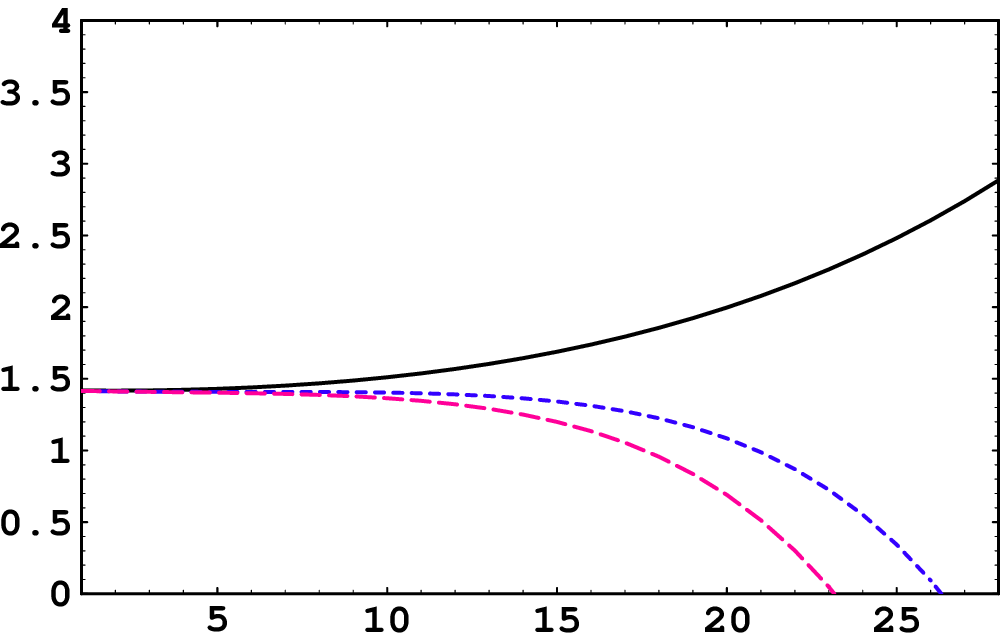}}
\put(100,-3){$m$}
\put(10,67){$F_0^A(0)$, GeV}
\end{picture}
\end{center}
\caption{The NRQCD sum rule results for the $F_{0}^A(0)$ form factor. The
contribution of $2S\to 2S$ transition has been taken into account. The
notations are the same as in Fig. \ref{nrfig1}.}
\label{nrfig2}
\end{figure}

\noindent
Further, in (\ref{nrhad}) we can take into account the dominant subleading
term, which is the contribution by the transition of $2S\to 2S$. In this case
one could expect that the form factor is not suppressed in comparison with the
contribution by the $1S\to 2S$ transition, since in the potential picture the
latter decay has to be neglected because the overlapping between the wave
functions at zero recoil is close to zero for the states with the different
quantum numbers \footnote{The corresponding estimates were performed in
\cite{cheng}, where the $1S\to 2S$ transition is suppressed with respect to
$1S\to 1S$ as 1/25.}. So, we can easily modify the relation (\ref{nrf0}) due to
the second transition and justify the value of $F_{0, 2S\to 2S}^{A}$ to reach
the stability of $F_{0, 1S\to 1S}^{A}$ at low values of moment numbers. We
find $F_{0, 2S\to 2S}^{A}/F_{0, 1S\to 1S}^{A} \approx 3.7$ and present the
behaviour of the form factor $F_0^A$ at zero recoil in Fig. \ref{nrfig2}.

\subsection{Contribution of the gluon condensate}

Following a general formalism for the calculation of gluon condensate
contribution in the Fock-Schwinger gauge \cite{Fock, Schw} we have considered
diagrams, depicted in Fig. \ref{cond}. In our calculations, we have used the
NRQCD approximation and analysed the limit, where the invariant mass of the
lepton pair takes its maximum value. Here we would like to note that 
the spin structure, in the leading order of relative velocity of heavy quarks, 
does not change, in comparison with the bare loop result for the
nonrelativistic quarks. Thus, we can conclude, that in this approximation, the
symmetries of NRQCD lagrangian lead to a universal Wilson coefficient for the
gluon condensate operator. As a consequence, relations (\ref{Fsym}) remain
valid.

Below we perform calculations for the form factor $F_{0}^{A}(p_1^2, p_2^2,
Q^2)$ in the limit $q^2\to q_{max}^2$. The contribution of gluon condensate to
the corresponding correlator is given by the following expression:
\begin{equation}
\Delta F_{0}^{G^2} =
\langle \frac{\alpha_s}{\pi}G_{\mu\nu}^2\rangle \cdot\frac{\pi}{48}[3R_{0} - 
R_{2}], \label{x1}
\end{equation}
where
\begin{eqnarray}
R_{0} &=& -\frac{1}{\pi^2i}\int\frac{k^2dkdk_0}{
P_3(k, k_0)P_1(k, k_0, \omega_1)P_2(k, k_0, \omega_2)}R_g ,\nonumber\\
R_{2} &=& -\frac{1}{\pi^2i}\int\frac{k^4dkdk_0}{
P_3(k, k_0)P_1(k, k_0, \omega_1)P_2(k, k_0, \omega_2)}R_k , \label{x2}
\end{eqnarray}
In these expressions the inverse propagators are
\begin{eqnarray}
P_1(k, k_0, \omega_1) &=& \omega_1 + k_0 - \frac{k^2}{2m_1}, \nonumber \\
P_2(k, k_0, \omega_2) &=& \omega_2 + k_0 - \frac{k^2}{2m_2},  \\
P_3(k, k_0) &=& - k_0 - \frac{k^2}{2m_3}, \nonumber 
\end{eqnarray}
The functions $R_g$ and $R_k$ are symmetric under the permutation of indices 1
and 2  and they have the following forms:
\begin{eqnarray}
R_g &=& -\frac{1}{m_1}\left [\frac{1}{P_3} + \frac{1}{P_1}\right
]\frac{1}{P_1^2} - 
\frac{1}{m_2}\left [\frac{1}{P_3} + \frac{1}{P_1}\right ]\frac{1}{P_2^2},
\nonumber\\
R_k &=& \frac{3}{m_1^2}\frac{1}{P_1^4} + \frac{3}{m_2^2}\frac{1}{P_2^4} +
\frac{3}{m_1^2}\frac{1}{P_1^3P_3} + \frac{3}{m_2^2}\frac{1}{P_2^3P_3} +
\nonumber \\
&& \frac{1}{m_1m_3}\frac{1}{P_1^2P_3^2} +
\frac{1}{m_2m_3}\frac{1}{P_2^2P_3^2} -
\frac{1}{m_3^2}\frac{1}{P_3^4} - \label{x3} \\
&&  \frac{1}{m_1m_3}\frac{1}{P_1P_3^3} - \frac{1}{m_2m_3}\frac{1}{P_2P_3^3}
-
\frac{1}{m_1m_2}\frac{1}{P_1P_2P_3^2} -
\frac{1}{m_1m_2}\frac{1}{P_1^2P_2^2}. \nonumber
\end{eqnarray}
Let us note, that in the calculations of diagrams in Fig. \ref{cond} we have
used the following vertex for the interaction of heavy quark with the gluon
\begin{equation}
L_{int}^{v} = -g_s\bar h_v v_{\mu}A^{\mu}h_v,
\end{equation}
where $A_{\mu} = A_{\mu}^{a}\cdot\frac{\lambda^{a}}{2}$ and its
Fourier-transform in Fock-Schwinger gauge has the form
\begin{equation}
A_{\mu}^{a}(k_g) = -\frac{i}{2}G_{\mu\nu}^{a}(0)\frac{\partial}{\partial k_g}
(2\pi)^4
\delta (k_g).
\end{equation}
After two differentiations of the nonrelativistic propagators
\begin{equation}
\frac{1}{P(p, m_Q)} = \frac{1}{pv + \frac{p^2}{2m_Q}},
\end{equation}
two types of contributions to the gluon condensate correction appear. The 
first is equal to
\begin{equation}
v_{\mu}v_{\nu}\alpha_s G_{\mu\alpha}^aG_{\nu\beta}^a\cdot g^{\alpha\beta}\to 
\langle \alpha_s G_{\mu\nu}^2\rangle \cdot\frac{3}{12},
\end{equation}
and it leads to the term with $R_{0}$. The second expression
\begin{equation}
v_{\mu}v_{\nu}\alpha_s G_{\mu\alpha}^aG_{\nu\beta}^ak^{\alpha}k^{\beta}\to 
\langle \alpha_sG_{\mu\nu}^2\rangle \frac{k^2 - (v\cdot k)^2}{12} = - \langle
\alpha_sG_{\mu\nu}^2\rangle 
\frac{|\vec k|^2}{12}
\end{equation}
leads to the term with $R_{2}$, which is of the same order in the relative
velocity of heavy quarks as $R_{0}$, but it is suppressed numerically in the
region of moderate numbers for the momenta of spectral densities, where the
non-perturbative contributions (condensates) of higher dimension operators are
not essential.

It is easy to show, that the contributions of $R_{0}$ and $R_{2}$ can be
obtained by the differentiation of two basic integrals:
\begin{eqnarray}
E_{0} &=& -\frac{1}{\pi^2i}\int_{0}^{\infty}\frac{k^2dkdk_0}{P_1P_2P_3} =
\frac{2m_{13}2m_{23}}{k_{13} + k_{23}}, \label{x4} \\
E_{2} &=& -\frac{1}{\pi^2i}\int_{0}^{\infty}\frac{k^4dkdk_0}{P_1P_2P_3} =
- \frac{2m_{13}2m_{23}}{k_{13} + k_{23}}(k_{13}^2 + k_{13}k_{23} +
k_{23}^2),\nonumber
\end{eqnarray}
where $k_{13} = \sqrt{-2m_{13}\omega_1}$, $k_{23} = \sqrt{-2m_{23}\omega_2}$.
Then for $R_{0}$ and $R_{2}$ we have:
\begin{eqnarray}
R_{0} &=& \hat R_{g}\cdot E_{0},\nonumber \\
R_{2} &=& \hat R_{k}\cdot E_{2},\nonumber 
\end{eqnarray}
where operators $\hat R_g$ and $\hat R_k$ can be obtained from $R_g$ and $R_k$
in (\ref{x3}) after the substitutions:
\begin{eqnarray}
\frac{1}{P_1^n}&\to& \frac{(-1)^n}{n!}
\frac{\partial^n}{\partial\omega_1^n}, \nonumber \\
\frac{1}{P_2^m}&\to& \frac{(-1)^m}{m!}
\frac{\partial^m}{\partial\omega_2^m}, \nonumber \\
\frac{1}{P_3^l}&\to& \frac{(-1)^l}{l!}
\left [\frac{\partial}{\partial\omega_1} +
\frac{\partial}{\partial\omega_2}\right ]^l, \nonumber
\end{eqnarray}
As a result we have:
\begin{eqnarray}
R_0 &=& \{\frac{1}{6m_1}\frac{\partial^3}{\partial\omega_1^3} +
\frac{1}{6m_2}\frac{\partial^3}{\partial\omega_2^3} + \frac{1}{2m_1}\left (
\frac{\partial^3}{\partial\omega_1^3} + 
\frac{\partial^3}{\partial\omega_1^2\partial\omega_2}\right ) + \nonumber\\
&& \frac{1}{2m_2}\left (\frac{\partial^3}{\partial\omega_2^3} + 
\frac{\partial^3}{\partial\omega_2^2\partial\omega_1}\right )\}\cdot
\frac{2m_{13}2m_{23}}{k_{13} + k_{23}}, \label{x5}\\
R_2 &=& \{\frac{3}{4!m_1^2}\frac{\partial^4}{\partial\omega_1^4} +
\frac{3}{4!m_2^2}\frac{\partial^4}{\partial\omega_2^4} + \frac{3}{3!m_1^2}\left
(
\frac{\partial^4}{\partial\omega_1^4} +
\frac{\partial^4}{\partial\omega_1^3\partial\omega_2}
\right ) + \nonumber \\
&& \frac{1}{3!m_2^2}\left (\frac{\partial^4}{\partial\omega_2^4} + 
\frac{\partial^4}{\partial\omega_2^3\partial\omega_1}\right ) +
\frac{1}{4m_1m_3}
\left (\frac{\partial^4}{\partial\omega_1^4} +
2\frac{\partial^4}{\partial\omega_1^3
\partial\omega_2} +
\frac{\partial^4}{\partial\omega_1^2\partial\omega_2^2}\right ) +
\nonumber \\
&& \frac{1}{4m_2m_3}\left (\frac{\partial^4}{\partial\omega_2^4} +
2\frac{\partial^4}{\partial\omega_2^3
\partial\omega_1} +
\frac{\partial^4}{\partial\omega_1^2\partial\omega_2^2}\right ) -
\nonumber \\
&& \frac{1}{4!m_3^2}\left (\frac{\partial^4}{\partial\omega_1^4} +
4\frac{\partial^4}{\partial\omega_1^3\partial\omega_2} + 
6\frac{\partial^4}{\partial\omega_1^2\partial\omega_2^2} + 
4\frac{\partial^4}{\partial\omega_1\partial\omega_2^3} +
\frac{\partial^4}{\partial\omega_2^4}
\right ) - \label{x6}\\ 
&& \frac{1}{3!m_1m_3}\left (\frac{\partial^4}{\partial\omega_1^4} +
3\frac{\partial^4}{\partial\omega_1^3\partial\omega_2} + 
3\frac{\partial^4}{\partial\omega_1^2\partial\omega_2^2} + 
\frac{\partial^4}{\partial\omega_1\partial\omega_2^3}\right ) - \nonumber\\ 
&& \frac{1}{3!m_2m_3}\left (\frac{\partial^4}{\partial\omega_2^4} +
3\frac{\partial^4}{\partial\omega_2^3\partial\omega_1} + 
3\frac{\partial^4}{\partial\omega_1^2\partial\omega_2^2} + 
\frac{\partial^4}{\partial\omega_2\partial\omega_1^3}\right ) - \nonumber\\
&& \frac{1}{2m_1m_2}\left (
\frac{\partial^4}{\partial\omega_1^3\partial\omega_2} + 
2\frac{\partial^4}{\partial\omega_1^2\partial\omega_2^2} + 
\frac{\partial^4}{\partial\omega_1\partial\omega_2^3}\right ) - 
\frac{1}{4m_1m_2}\left (
\frac{\partial^4}{\partial\omega_1^2\partial\omega_2^2}\right )\}\cdot 
\nonumber\\ 
&& 
\left (-\frac{2m_{13}2m_{23}}{m_{13} + m_{23}}\right )(k_{13}^2 + k_{13}k_{23}
+ k_{23}^2),
\nonumber
\end{eqnarray}
Equations (\ref{x1}), (\ref{x5}) and (\ref{x6}) represent the most compact
analytical expression for the contribution of gluon condensate to the
form factor $F_0^A$, whereas performing the differentiations leads to very
cumbersome expressions.

In the moment scheme of sum rules we suppose
\begin{eqnarray}
\omega_1 &=& -(m_1 + m_3) + q_1, \nonumber\\
\omega_2 &=& -(m_2 + m_3) + q_2, \nonumber
\end{eqnarray}
and expand $\Delta F_{0}^{G^2}$ in a series over $\{q_1, q_2\}$ at the point
$\{ 0, 0\}$, which allows us to determine $\Delta F_{0}^{G^2}[n, m] =
\frac{1}{n!} \frac{1}{m!}\frac{d^{n+m}}{dq_1^ndq_2^m}\Delta F_{0}^{G^2}$.
Further analysis has been performed numerically with the help of MATHEMATICA.
The evaluation of the $\Delta F_0^{G^2}[n,m]$ dependence in a broad range of
$[n,m]$ takes too much calculation time, so we restrict ourselves by showing
the results in Figs. \ref{nrfig1} and \ref{nrfig2} for the fixed $n=4$ with the
following set of parameters
\begin{eqnarray}
m_b &=& 4.6\; \mbox{GeV},\nonumber\\
m_c &=& 1.4\; \mbox{GeV},\nonumber\\
\langle\frac{\alpha_s}{\pi}G_{\mu\nu}^2\rangle &=& 1.7\cdot 10^{-2}\;
\mbox{GeV}^4.\nonumber 
\end{eqnarray}
by taking into  
account (Fig. \ref{nrfig1}) and without accounting (Fig. \ref{nrfig2}) 
the contribution of transition $2S\to 2S$, where $F_{0}(2S\to 2S)/F_{0}(1S\to
1S)\approx 3.7$. As can be seen in these figures, the gluon
contribution, while varying the moment number in the region of bound $\bar
cc$-states, plays an important role, because it allows us to extend the
stability region for the form factor $F_{0}^A$ up to three times (from $m < 5$
till $m < 15$), and, thus, the reliability of sum rule predictions. Let us also
note, that in the scheme of saturation for the hadronic part of sum rules by
the ground states in both variables $s_1$ and $s_2$, the account for the gluon
condensate leads to the $20\%$ reduction for the value of form factor
$F_{0}^{A}$.

The analysis of the dependence on the moment number $m$ in the region of bound
states $\bar b c$ at fixed $m$ shows that the contribution of gluon condensate
in this case does not affect the character of this dependence. This may be
explained by the fact that the Coulomb corrections to the Wilson coefficient
of gluon operator $G_{\mu\nu}^2$ play an essential role\footnote{
At present, there is an analytical calculation of initial six moments for the
Wilson coefficient of gluon condensate in the second order of $\alpha_s$
\cite{iliyn}. In this region, the influence of gluon condensate to the sum rule
results is negligibly small for the heavy quarkonia, containing the $b$-quark,
which does not allow one to draw definite conclusions on the role of such
$\alpha_s$-corrections.}. The summation of
$\alpha_s/v$-terms for $\bar bc$ may give a sizeable effect, contrary to the
situation with the $\bar cc$ system, where the relative velocity of heavy
quarks is not too small.

To conclude, we have calculated the contribution of gluon condensate to the
three-point sum rules for heavy quarkonia in the leading order of relative
velocity of heavy quarks and in the first order of $\alpha_s$. Due to the
symmetry of NRQCD in the limit, when the invariant mass of the lepton pair
takes its maximum value, i.e. at the recoil momentum close to zero, the Wilson
coefficient for the form factor $F_{0}^A$ is universal in the sense that it
determines the contributions of gluon condensate to other form factors, in
accordance with relations (\ref{Fsym}).

\section{Conclusions}

\noindent
We have calculated the semileptonic decays of $B_c$ meson in the framework of
sum rules in QCD and NRQCD. We have extended the previous evaluations in QCD to
the case of massive leptons: the complete set of double spectral densities in
the bare quark-loop approximation have been presented. The analysis in the sum
rule schemes of density moments and Borel transform has been performed and 
consistent results have been obtained.

We have taken into account the gluon condensate contribution for the
form factors of semipeltonic transitions between the heavy quarkonia in the
Borel transform sum rules of QCD, wherein the analitycal expressions for the
case of three nonzero masses of quarks have been presented.

We have considered the soft limit on the form factors in NRQCD at the recoil
momentum close to zero, which has allowed one to derive the generalized
relations due to the spin symmetry of effective lagrangian. The relations have
shown a good agreement with the numerical estimates in full QCD, which means
the corrections in both the relative velocities of heavy quarks inside the
quarkonia and the inverse heavy quark masses to be small within the accuracy
of the method. Next, we have presented the analytical results on the gluon
condensate term in the NRQCD sum rules within the moments scheme.

In both the QCD and NRQCD sum rules, the account for the gluon condensate has
allowed one to enforce the reliability of predictions, since the region of
physical stability for the form factors evaluated has significantly expanded in
comparison with the leading order calculations of bare quark-loop contribution.

Next, we have investigated the role played by the Coulomb $\alpha_s/v$
corrections for the semileptonic transitions between the heavy quarkonia.
We have shown that as in the case of two-point sum rules, the three-point
spectral densities are enhanced due to the Coulomb renormalization of 
quark-meson vertices. 

The complete analysis shows that the numerical estimates of various branching
fractions:
\begin{eqnarray}
{\rm BR}[B_c^+\to J/\psi l^+ \nu]  &=&  2.5\pm 0.5 \% ,\nonumber\\
{\rm BR}[B_c^+\to \bar c c+\mbox{\it X}]  &=&  23\pm 5 \% , \nonumber
\end{eqnarray}
agree with the results obtained in the framework of potential models and
Operator Product Expansion in NRQCD. More detailed results are presented in
tables.

Thus, we draw the conclusion that at present the theoretical predictions on the
semileptonic decays of $B_c$ meson give consistent and reliable results.

The authors would like to express the gratitude to Prof. A.Wagner and members
of DESY Theory Group for their kind hospitality during the visit to DESY, where
this paper was written, as well as to Prof. S.S.Gershtein for discussions. The
authors also thank Prof. A. Ali for reading this manuscript and valuable
remarks. V.V.K. and A.K.L. express the gratitude to A.L.Kataev for fruitful
discussions and notes. 

This work is in part supported by the Russian Foundation for Basic Research,
grants 99-02-16558 and 96-15-96575. The work of A.I.Onishchenko was supported,
in part, by International Center of Fundamental Physics in Moscow, Intenational
Science Foundation, and INTAS-RFBR-95I1300 grants.

\section{Appendix A}

In this Appendix we present the derivation of exclusive semileptonic widths for
the $B_c$-meson decays into the $J/\psi$, $\eta_c$ mesons with account of
lepton masses. 

The exclusive semileptonic width $\Gamma_{SL}$ for the decay 
$B_c\to J/\psi (\eta_c) l\bar\nu_l$, where $l = e,\mu$ or $\tau$, can be
written down in the form \cite{Bigi}
\begin{equation}
\Gamma_{SL} = \frac{1}{(2\pi)^3}\frac{G_F^2|V_{cb}|^2}{M_{B_c}}\int d^4q
\int d\tau_l L^{\alpha\beta}W_{\alpha\beta}, \label{SL1}
\end{equation}
where $d^4q = 2\pi |\vec q|dq^2dq_0$, $d\tau_l = |\vec p_l|d\Omega_l
/(16\pi^2\sqrt{q^2})$ is the leptonic pair phase space, $d\Omega_l$ is the 
solid angle of charged lepton $l$, $|\vec p_l| = \sqrt{q^2}\Phi_l/2$ is its 
momentum in the dilepton center of mass system and $\Phi_l\equiv\sqrt{1 - 
2\lambda_{+} + \lambda_{-}^2}$, with $\lambda_{\pm}\equiv 
(m_l^2\pm m_{\nu_l}^2)/q^2$. The tensors $L^{\alpha\beta}$ and 
$W_{\alpha\beta}$ in Eq. (\ref{SL1}) are given by
\begin{equation}
L^{\alpha\beta} = \frac{1}{4}\sum_{spins} (\bar lO^{\alpha}\nu_l)
(\bar\nu_l O^{\beta}l) = 2[p_l^{\alpha}p_{\nu_l}^{\beta} + p_l^{\beta}
p_{\nu_l}^{\alpha} - g^{\alpha\beta}(p_l\cdot p_{\nu_l}) + 
i\epsilon^{\alpha\beta\gamma\delta}p_{l\gamma}p_{\nu_l\delta}], \label{SL2}
\end{equation} 
and
\begin{eqnarray}
W_{\alpha\beta} &=& \int\frac{d^3\vec p_2}{2E_2}\delta^4(p_1 - p_2 - q)\tilde
W_{\alpha\beta}\nonumber \\
&=& \theta (E_2)\delta (M_1^2 - 2M_1\cdot q_0 + q^2 - M_2^2)
\tilde W_{\alpha\beta}|_{p_2 = p_1 - q},\label{SL3}
\end{eqnarray}
where 
\begin{equation}
\tilde W_{\alpha\beta} = (f_{+}(t)(p_1 + p_2)_{\alpha} + f_{-}(t)q_{\alpha})
(f_{+}(t)(p_1 + p_2)_{\beta} + f_{-}(t)q_{\beta}) 
\end{equation}
for the pseudoscalar particle in the final state, and
\begin{eqnarray}
\tilde W_{\alpha\beta}& =& -(iF_{0}^{A}g_{\alpha e} + 
iF_{+}^{A}p_{1e}(p_1 + p_2)_{\alpha} + iF_{-}^{A}p_{1e}q_{\alpha} - 
F_V\epsilon_{\alpha eij}(p_1 + p_2)^iq^j)\cdot \nonumber\\
&& (iF_{0}^{A} + iF_{+}^{A}p_{1k}(p_1 + p_2)_{\beta} + 
iF_{-}^{A}p_{1k}q_{\beta} + F_V\epsilon_{\beta kmn}(p_1 + p_2)^mq^n )\cdot
\nonumber\\
&& (\sum_{polarizations}\epsilon^{e}\epsilon^{*k}) 
\end{eqnarray}
for the vector meson in the final state with the polarization $\epsilon_{\mu}$, 
where 
\begin{equation}
\sum_{polarizations}\epsilon^{e}\epsilon^{*k} = \frac{p_{2}^{e}p_{2}^{k}}
{M_2^2} - g^{ek}
\end{equation}
$M_2$ is the mass of final state meson ($J/\psi$ or $\eta_c$). 

The integral over the leptonic phase space in Eq.(\ref{SL1}) is given by
\begin{equation}
\int d\tau_l L^{\alpha\beta} = \frac{1}{4\pi}\frac{|\vec p_l|}{\sqrt{q^2}}
\langle L^{\alpha\beta}\rangle,
\end{equation}
with
\begin{equation}
\langle L^{\alpha\beta}\rangle = \frac{1}{4\pi}\int d\Omega_l 
L^{\alpha\beta} = \frac{2}{3}\{ (1 + \lambda_1)(q^{\alpha}q^{\beta} - 
g^{\alpha\beta}q^2) + \frac{3}{2}\lambda_2 g^{\alpha\beta}q^2\},
\end{equation}
where $\lambda_1\equiv \lambda_{+} - 2\lambda_{-}^2$ and $\lambda_2\equiv 
\lambda_{+} - \lambda_{-}^2$. Introducing the dimensionless kinematical
variable $t\equiv q^2/m_b^2$ and integrating over $q_0$ the semileptonic 
width takes the following form:
\begin{equation}
\Gamma_{SL} = \frac{1}{64\pi^3}\frac{G_F^2|V_{bc}|^2m_b^2}{M_1^2}
\int_{t_{min}}^{t_{max}} dt\Phi_l(t)|\vec q|\langle L^{\alpha\beta}\rangle
\tilde W_{\alpha\beta}, \label{SL4}
\end{equation}
where 
$$
|\vec q| = \frac{1}{2M_1}\sqrt{(M_1^2 + m_b^2t - M_2^2)^2 - 
4m_b^2M_1^2t}.
$$
In Eq. (\ref{SL4}) the limits of integration are given by $t_{min} = 
\frac{m_l^2}{m_b^2}$ and $t_{max} = \frac{1}{m_b^2}(M_1 - M_2)^2$.

Calculation of $\langle L^{\alpha\beta}\rangle\tilde  W_{\alpha\beta}$ yields
the following expressions
\begin{eqnarray}
\langle L^{\alpha\beta}\rangle\tilde  W_{\alpha\beta} &=& \frac{1}{3}(
3t^2f_{-}^2(t)\lambda_2 m_b^4 + 6tf_{-}(t)f_{+}(t)(M_1^2-M_2^2)
\lambda_2 m_b^2 +\nonumber\\ 
&& f_{+}^2(t)(t^2(2\lambda_1 - 3\lambda_2 + 2)m_b^4 - 
2t(M_{1}^2 + M_{2}^2)(2\lambda_1 - 3\lambda_2 + 2)m_b^2 +\nonumber \\
&& 2(M_{1}^2 - M_{2}^2)^2(\lambda_1 + 1))) \nonumber
\end{eqnarray}
for the pseudoscalar meson in the final state, and
\begin{eqnarray}
\langle L^{\alpha\beta}\rangle\tilde  W_{\alpha\beta} &=& \frac{1}{12M_{2}}
(2(t^2(\lambda_1+1)m_b^4-2t((1+\lambda_1)M_{1}^2 - 5(1+\lambda_1)M_{2}^2 +
9M_{2}^2\lambda_2)m_b^2+\nonumber\\
&& (M_{1}^2-M_{2}^2)^2(1+\lambda_1))(F_{0}^{A})^2 -
2(t^2m_b^4 - 2t(M_{1}^2+M_{2}^2)m_b^2 + \nonumber \\
&& (M_{1}^2-M_{2}^2)^2)
(F_{+}^{A}(2(1+\lambda_1)tm_b^2-3t\lambda_2m_b^2-2(1+\lambda_1)M_{1}^2+
\nonumber\\
&& 2(1+\lambda_1)M_{2}^2) - 3tF_{-}^{A}m_b^2\lambda_2)F_{0}^{A} +
(t^2m_b^4-2t(M_{1}^2+M_{2}^2)m_b^2+\nonumber \\
&& (M_{1}^2-M_{2}^2)^2)((2t^2m_b^4+
2t^2\lambda_1m_b^4 - 3t^2\lambda_2m_b^4 - 4(1+\lambda_1)tM_{1}^2m_b^2 -
\nonumber\\
&& 4(1+\lambda_1)tM_{2}^2 + 6tM_{1}^2\lambda_2m_b^2 + 6tM_{2}^2\lambda_2m_b^2
+ 2(1+\lambda_1)M_{1}^4 +\nonumber\\
&& 2(1+\lambda_1)M_{2}^4 - 4(1+\lambda_1)M_{1}^2M_{2}^2)(F_{+}^{A})^2 + 
\nonumber\\
&& 6tF_{-}^{A}F_{+}^{A}\lambda_2m_b^2(M_{1}^2-M_{2}^2) +
tm_b^2(3t(F_{-}^{A})^2\lambda_2m_b^2 + \nonumber\\
&& 16(1+\lambda_1)(F_{V})^2M_{2}^2 -
24\lambda_2M_{2}^2(F_{V})^2))) \nonumber
\end{eqnarray}
for the vector meson in the final state.

\section{Appendix B}

In this Appendix we illustrate the kind of expressions, which arise
for the gluon condensate contribution to the form factors $F^{i}(t)$ in the
framework of Borel transformed three point sum rules, in the case of $f_1(t) =
f_{+}(t) + f_{-}(t)$ form factor.

\noindent
Following the algorithm, described in section {\bf 2.3}, for 
$\Pi_1^{\langle G^2\rangle}$ we have the following expression
\begin{eqnarray}
\Pi_1^{\langle G^2\rangle} &=& C_1^{(-1,-2)}U_0(-1,-2) +
C_1^{(-1,-1)}U_0(-1,-1) +
\sum_{i=-4}^{0} C_1^{(0,i)}U_0(0,i) +\nonumber\\ 
&& \sum_{i=-5}^{-1} C_1^{(1,i)}U_0(1,i) + 
 \sum_{i=-6}^{0} C_1^{(2,i)}U_0(2,i) + \sum_{i=-6}^{0} C_1^{(3,i)}U_0(3,i),
\end{eqnarray}  
where
\begin{eqnarray}
C_1^{(-1,-2)} &=& -\frac{M_{bc}^2+M_{cc}^2}{12M_{bc}^2}, \\
C_1^{(-1,-1)} &=& -\frac{1}{12M_{bc}^2},\\
C_1^{(0,-4)} &=& \frac{1}{48M_{bc}^6M_{cc}^4}(M_{bc}^{10}(6m_2^2+m_3^2) +
8m_2^2M_{bc}^8M_{cc}^2 + (-5m_1^2+5m_2^2+\nonumber\\
&& m_3^2)M_{bc}^6M_{cc}^4
+ (-7m_1^2+3m_2^2+2m_1(m_2-m_3))M_{bc}^4M_{cc}^6 - 4m_1^2M_{bc}^2M_{cc}^8 
\quad\quad\quad\\
&& - 2m_1^2M_{cc}^{10}),\nonumber\\
C_1^{(0,-3)} &=& \frac{1}{48M_{bc}^6M_{cc}^4}((m_1^2-2m_1m_3-2m_3^2)M_{cc}^8
+ (6m_2^2+4M_{cc}^2)M_{bc}^8 - \nonumber\\
&& (2m_1^2+2m_2^2+6m_2m_3+2m_3^2+2m_1(-5m_2+m_3)-
11M_{cc}^2\nonumber\\
&& -2Q^2)M_{bc}^6M_{cc}^2 + (m_1^2-m_2^2+2m_1(m_2-5m_3)-2m_2m_3+2m_3^2 
\\
&& + 5M_{cc}^2+Q^2)M_{bc}^2M_{cc}^6 + (-7m_1^2+4m_1m_2-2m_2^2-10m_1m_3-
\nonumber\\
&& 8m_2m_3+2m_3^2+12M_{cc}^2+2Q^2)M_{bc}^4M_{cc}^4),\nonumber\\
C_1^{(0,-2)} &=& \frac{1}{48M_{bc}^6M_{cc}^4}((3m_1^2-2m_1m_3-2m_3^2)M_{cc}^6
+ (m_2^2-5m_3^2+9M_{cc}^2)M_{bc}^6 +\nonumber\\
&& (-4m_1^2-7m_2^2+2m_1(6m_2-5m_3)-8m_2m_3-
13m_3^2+16M_{cc}^2+\nonumber\\
&& 4Q^2)M_{bc}^4M_{cc}^2 - (5m_1^2-4m_1(m_2-4m_3)+2(m_2^2+
2m_2m_3+m_3^2-\\
&& M_{cc}^2-Q^2)M_{bc}^2M_{cc}^4),\nonumber\\
C_1^{(0,-1)} &=& \frac{1}{48M_{bc}^6M_{cc}^4}(-2m_3^2M_{cc}^4 + 
(-6m_3^2+5M_{cc}^2)M_{bc}^4 + \\
&&
(-m_1^2+2m_1m_2-m_2^2+4m_2m_3-16m_3^2+M_{cc}^2+Q^2)M_{bc}^2M_{cc}^2,\nonumber\\
C_1^{(0,0)} &=& \frac{m_3^2(M_{bc}^2+M_{cc}^2)}{48M_{bc}^6M_{cc}^4},\\
C_1^{(1,-5)} &=& -\frac{1}{48M_{bc}^8M_{cc}^6}(m_2^2M_{bc}^4-m_1^2M_{cc}^4)
((5m_2^2+m_3^2)M_{bc}^8 - m_3^2M_{bc}^6M_{cc}^2 - \\
&& (4m_1^2+3m_2^2)M_{bc}^4M_{cc}^4
+ 2m_1(m_2-m_3)M_{bc}^2M_{cc}^6 + 2m_1^2M_{cc}^4) ,\nonumber\\
C_1^{(1,-4)} &=& \frac{1}{48M_{bc}^8M_{cc}^6}(m_1^2(3m_1^2+2m_1m_3+6m_3^2)
M_{cc}^{10} + \nonumber\\
&& (-2m_2^4-7m_2^2M_{cc}^2+m_3^2M_{cc}^2)M_{bc}^{10} +
(8m_2^4+4m_2^3m_3+3m_2^2m_3^2+\nonumber\\
&& 2m_2m_3^3+m_1^2(4m_2^2+m_3^2)-2m_1(4m_2^3-3m_2^2m_3+
m_2m_3^2-m_3^3- \\
&& 11m_2^2M_{cc}^2-3m_3^2M_{cc}^2-4m_2^2Q^2-m_3^2Q^2)M_{bc}^8M_{cc}^2
- (3m_1^4+6m_2^2m_3^2+\nonumber\\
&& m_1^3(-6m_2+4m_3)+2m_1(m_2^2m_3+2m_2M_{cc}^2-2m_3M_{cc}^2) +\nonumber\\
&& m_1^2(10m_2^2+2m_2m_3+2m_3^2-8M_{cc}^2-3Q^2))M_{bc}^4M_{cc}^6+ 
(4m_1m_2^2m_3+\nonumber\\
&& m_1^2(9m_2^2+16M_{cc}^2)+m_2^2(2m_2^2+4m_2M-3-4m_3^2-M_{cc}^2-\nonumber\\
&& 2Q^2))M_{bc}^6M_{cc}^4-m_1(6m_1^3+4m_1^2m_3+2m_3^2(-m_2+m_3)+
m_1(m_2^2+\nonumber\\
&& 2m_2m_3-4m_3^2+5M_{cc}^2-Q^2))M_{bc}^2M_{cc}^8),\nonumber\\
C_1^{(1,-3)} &=& \frac{1}{48M_{bc}^{8}M_{cc}^6}(2m_3(m_1^3+5m_1^2m_3+2m_3^3)
M_{cc}^8 + (m_3^4+m_2^2(9m_3^2-4M_{cc}^2+\nonumber\\
&& 8m_2m_3M_{cc}^2-3m_3^2M_{cc}^2-11M_{cc}^4)
M_{bc}^8 + (2m_2^4+10m_2^2m_3^2+m_3^4+\nonumber\\
&& 7m_2^2M_{cc}^2+18m_2m_3M_{cc}^2+
2m_3^2M_{cc}^2-20M_{cc}^2-4m_1(m_2-m_3)(m_2^2\\
&& +2M_{cc}^2)+m_1^2(2m_2^2+5M_{cc}^2)-
2m_2^2Q^2-5M_{cc}^2Q^2)M_{bc}^6M_{cc}^2 + (-7m_2^2m_3^2\nonumber\\
&& -4m_3^4+4m_2^2M_{cc}^2+10m_2m_3M_{cc}^2-7M_{cc}^4+2m_1^2(2m_2m_3-
m_3^2+5M_{cc}^2)+\nonumber\\
&&m_1(6m_2-6m_3-
2m_2^2m_3+4m_2m_3^2+4m_2M_{cc}^2+10m_3M_{cc}^2)+2m_3^2Q^2\nonumber\\
&& -4M_{cc}^2Q^2)M_{bc}^4
M_{cc}^6 + (m_1^4-2m_1^3(m_2-4m_3)+10m_3^2M_{cc}^2+2m_1m_3(m_2^2+\nonumber\\
&& m_2m_3+m_3^2-M_{cc}^2-Q^2)+m_1^2(m_2^2+2m_2m_3+7M_{cc}^2-Q^2)),\nonumber\\
C_1^{(1,-2)} &=& \frac{1}{48M_{bc}^8M_{cc}^6}(m_3^2(3m_1^2-2m_1m_3+6m_3^2)
M_{cc}^6 + (14m_2m_3M_{cc}^2+(16m_3^2-\nonumber\\
&& 13M_{cc}^2)M_{cc}^2+m_2^2(4m_3^2+M_{cc}^2))
M_{bc}^6 + (-5m_2^2m_3^2-2m_2m_3^3-2m_3^4+\nonumber\\
&& 4m_2^2M_{cc}^2+8m_2m_3M_{cc}^2+
32m_3^2M_{cc}^2-8M_{cc}^4+m_1^2(-3m_3^2+4M_{cc}^2)+\\
&& 2m_1(3m_2m_3^2-2m_3^3-2m_2M_{cc}^2
+6m_3M_{cc}^2)+3m_3^2Q^2-4M_{cc}^2Q^2)M_{bc}^4M_{cc}^2\nonumber\\
&&-(2m_1^3m_3+m_1^2(-4m_2m_3+4m_3^2+M_{cc}^2)+m_3^2(m_2^2+2m_2m_3+
4m_3^2\nonumber\\
&& -7M_{cc}^2-Q^2)+2m_1m_3
(m_2^2-m_2m_3+3m_3^2-2M_{cc}^2-Q^2))M_{bc}^2M_{cc}^6),\nonumber\\
C_1^{(1,-1)} &=& -\frac{1}{48M_{bc}^8M_{cc}^6}(2m_1m_3^3M_{cc}^4 +
(4m_3^4-17m_3^2M_{cc}^2+M_{cc}^4)M_{bc}^4 +\nonumber\\
&& 2m_3^2(m_1^2-2m_1m_2+m_2^2+2m_1m_3+4m_3^2-6M_{cc}^2-Q^2)),\\
C_1^{(2,-6)} &=& \frac{(M_{bc}^2-M_{cc}^2)(m_2^2M_{bc}^4-m_1^2M_{cc}^4)^3}
{96M_{bc}^{10}M_{cc}^8},\\
C_1^{(2,-5)} &=& -\frac{1}{96M_{bc}^8M_{cc}^6}(m_2^2M_{bc}^4-m_1^2M_{cc}^4)
((m_2^2+2m_3^2)M_{bc}^6 + 7m_1^2M_{bc}^2M_{cc}^4 +\nonumber\\
&&  m_2^2(m_1^2-2m_1m_2+m_2^2+
2m_1m_3+2m_2m_3+3M_{cc}^2-Q^2)M_{bc}^4 - \\
&&  m_1^2(m_1^2-2m_1m_2+m_2^2+2m_1m_3+2m_2m_3+
7M_{cc}^2-Q^2)M_{cc}^4),\nonumber\\
C_1^{(2,-4)} &=& \frac{1}{96M_{bc}^{10}M_{cc}^8}(-3m_1^4m_3^2M_{cc}^{10}+
m_1^3(6m_2M_{bc}^2-6m_3M_{bc}^2+m_1m_3^2-\nonumber\\
&& 2m_1M_{cc}^2-4m_3M_{cc}^2)M_{bc}^2M_{cc}^8 -
(12m_2^3m_3M_{cc}^2+4m_2m_3^3M_{cc}^2-12m_2^2M_{cc}^4\nonumber\\
&& -4m_3^2M_{cc}^4+m_2^4(3m_3^2+4M_{cc}^2))M_{bc}^{10} +
2m_1(-3m_2^3M_{bc}^2+4m_1m_2m_3M_{cc}^2\nonumber\\
&&-8m_1M_{cc}^4+m_2^2(3m_3M_{bc}^2-m_1m_3^2+3m_1M_{cc}^2+2m_3M_{cc}^2))M_{bc}^6
M_{cc}^4 + \nonumber\\
&& 2m_1^2(-4m_2(m_1-2m_3)M_{cc}^2+m_2^2(m_3^2+3M_{cc}^2)+(3m_1^2+6m_1m_3+
\\
&& 2M_{cc}^2-3Q^2)M_{cc}^2)M_{bc}^4M_{cc}^6 + m_2^2(m_2^2(m_3^2-4M_{cc}^2)+
4m_2(m_1-4m_3)M_{cc}^2\nonumber\\
&& +2(-2m_1^2-4m_1m_3+3M_{cc}^2+2Q^2)M_{cc}^2)M_{bc}^8M_{cc}^2,\nonumber\\
C_1^{(2,-3)} &=& \frac{1}{48M_{bc}^8M_{cc}^6}(m_1(9m_3M_{bc}^2-m_1m_3^2+
m_2(-9M_{bc}^2-4m_1m_3+4m_3^2)+\nonumber\\
&& 4m_1M_{cc}^2)M_{bc}^2M_{cc}^4 - (2m_2^3m_3+
9m_2^2m_3^2+m_3^4+10m_2m_3M_{cc}^2-\\
&& 10M_{cc}^4)M_{bc}^6 + (m_2^4m_3^2+2m_2^3m_3^3-
3m_2^2m_3^2M_{cc}^2-3m_2^2M_{cc}^4-14m_2m_3M_{cc}^4\nonumber\\
&& +9M_{cc}^6+m_1^2(m_2^2m_3^2+
2m_3^2M_{cc}^2-3M_{cc}^4)+2m_1(-m_2^3m_3^2+m_2^2m_3^3+\nonumber\\
&& m_2^2m_3M_{cc}^2-
2m_2m_3^2M_{cc}^2+m_2M_{cc}^4-3m_3M_{cc}^4)-m_2^2m_3^2Q^2-2m_3^2M_{cc}^2Q^2+
\nonumber\\
&& 3M_{cc}^2Q^2)M_{bc}^4+m_1(m_1^3m_3^2+3(m_2-m_3)m_3^2M_{bc}^2+
m_1^2(3m_2M_{bc}^2-3m_3M_{bc}^2\nonumber\\
&& -2m_2m_3^2+2m_3^3-2m_3M_{cc}^2)+
m_1m_3^2(m_2^2+2m_2m_3-3M_{cc}^2-Q^2))M_{cc}^4),\nonumber\\
C_1^{(2,-2)} &=& \frac{1}{96M_{bc}^{10}M_{cc}^8}(3m_1^2m_3^4M_{cc}^6+
m_1m_3^2M_{bc}^2M_{cc}^4(6m_2M_{bc}^2-6m_3M_{bc}^2+m_1m_3^2-\nonumber\\
&& 2m_1M_{cc}^2+
4m_3M_{cc}^2)+M_{bc}^6(3m_2^2m_3^4+8m_2m_3(m_3^2-M_{cc}^2)M_{cc}^2+\\
&& 4M_{cc}^4
(-11m_3^2+2M_{cc}^2)) +
M_{bc}^4m_3M_{cc}^2(-8m_2(m_1-2m_3)m_3M_{cc}^2+\nonumber\\
&&
m_2^2(m_3^3+6m_3M_{cc}^2)-2M_{cc}^2(-3m_1^2m_3-6m_1m_3^2+2m_1M_{cc}^2+\nonumber
\\
&& 20m_3M_{cc}^2+3m_3Q^2))),\nonumber\\
C_1^{(2,-1)} &=& \frac{1}{96M_{bc}^8M_{cc}^6}m_3^2(M_{bc}^2(4m_2m_3+15m_3^2-
16M_{cc}^2)+m_3(-m_1^2m_3+\\
&& 2m_1(m_2m_3-m_3^2+2M_{cc}^2)+m_3(-m_2^2-2m_2m_3+13M_{cc}^2
+Q^2))),\nonumber\\
C_1^{(2,0)} &=& -\frac{m_3^4(M_{bc}^2(m_3^2-4M_{cc}^2)+m_3^2M_{cc}^2}
{96M_{bc}^{10}M_{cc}^8},\\
C_1^{(3,-6)} &=& \frac{(M_{bc}^4m_2^2-m_1^2M_{cc}^2)^3}
{96M_{bc}^{10}M_{cc}^8},\\
C_1^{(3,-5)} &=& \frac{1}{96M_{bc}^8M_{cc}^8}(M_{bc}^4m_2^2-m_1^2M_{cc}^4)
(M_{bc}^4m_2^2(2m_2m_3+M_{cc}^2) -\nonumber\\
&&  m_1^2M_{cc}^4(2m_2m_3+7M_{cc}^2)),\\
C_1^{(3,-4)} &=& -\frac{1}{96M_{bc}^{10}M_{cc}^8}(-3m_1^4m_3^2M_{cc}^8 +
M_{bc}^8m_2^2(m_2^2m_3^2-8m_2m_3M_{cc}+2M_{cc}^4)+\nonumber\\
&& 2M_{bc}^4m_1^2M_{cc}^2(m_2^2m_3^2+6m_2m_3M_{cc}^2+3M_{cc}^4)),\\
C_1^{(3,-3)} &=& \frac{1}{48M_{bc}^8M_{cc}^8}(m_1^2m_3^2M_{cc}^4
(-2m_2m_3+M_{cc}^2) + M_{bc}^4(-2m_2^3m_3^3+2m_2^2m_3^2M_{cc}^2+\nonumber\\
&& 6m_2m_3M_{cc}^4-3M_{cc}^6)),\\
C_1^{(3,-2)} &=& -\frac{m_3^2(3m_1^2m_3^2M_{cc}^4 + M_{bc}^4(m_2^2m_3^2+
12m_2m_3M_{cc}^2-18M_{cc}^4))}{96M_{bc}^{10}M_{cc}^8},\\
C_1^{(3,-1)} &=& \frac{m_3^4(2m_2M-3-9M_{cc}^2)}{96M_{bc}^8M_{cc}^8},\\
C_1^{(3,0)} &=& \frac{m_3^6}{96M_{bc}^{10}M_{cc}^8}.
\end{eqnarray}

For functions $U_0(a, b)$ we have the following expressions
$$
U_0(a,b) = \sum_{n = 1+b}^{1+a+b}2C_{n-b-1}^{a}\exp[-B_0]
(M_{bc}^2 + M_{cc}^2)^{a+b+1-n}\left (\frac{B_{-1}}{B_{1}}\right )^{\frac{n}
{2}}K_{-n} [2\sqrt{B_{-1}B_{1}}], \nonumber
$$
for $a\geq 0$. Here $K_{n}[z]$ is the modified Bessel function of the second 
order. In the case of $U_0(-1,-2)$ and $U_0(-1,-1)$ we have failed to obtain
exact analytical expressions, so we present their analytical approximations:
\begin{eqnarray}
U_0(-1,-2) &=& \frac{\exp [-B_0]}{2v^3B_{-1}B_{1}}\{ 2
\exp[-\sqrt{\frac{B_{-1}}{B_{1}}}]B_{1}v^2 + 2\exp [-B_{-1}^{1/2}B_{1}^{3/2}]
\sqrt{B_{-1}B_{1}}v^2\nonumber\\
&& + \exp [-B_{-1}^{1/2}B_{1}^{3/2}]B_{-1}^{3/2}B_{1}^{1/2}(2+vB_{1})v -
\nonumber\\
&& 2v^2B_{-1}B_{1}^2\left (\exp [\frac{B_{-1}}{v}]\Gamma \left (0, 
\frac{B_{-1}}{v} + \sqrt{\frac{B_{-1}}{B_{1}}} \right )
+ \Gamma (0, B_{-1}^{1/2}B_{1}^{3/2})\right )\nonumber\\
&& -2vB_{-1}B_{1}(\exp [\frac{B_{-1}}{v}]\Gamma\left (0, \frac{B_{-1}}{v}
+ \sqrt{\frac{B_{-1}}{B_{1}}} \right) + \Gamma (0, B_{-1}^{1/2}B_{1}^{3/2})
\nonumber\\ 
&& - \exp[vB_{1}]\Gamma (0, B_{-1}^{1/2}B_{1}^{3/2} + vB_{1})) - 
v^2B_{-1}\exp [B_{-1}^{1/2}B_{1}^{3/2}]\\
&& - B_{-1}^2B_{1}(v^2B_{1}^2\Gamma (0, B_{-1}^{1/2}B_{1}^{3/2}) +
2vB_{1}\Gamma (0, B_{-1}^{1/2}B_{1}^{3/2}) \nonumber\\
&& + 2\Gamma (0, B_{-1}^{1/2}
B_{1}^{3/2} - 2\exp [vB_{1}]\Gamma (0, B_{-1}^{1/2}B_{1}^{3/2} + vB_{1}))\},
\nonumber\\
U_0 (-1, -1) &=& \frac{-\exp[B_0]}{v^2}\{ (\exp\frac{B_{-1}}{v}
\Gamma \left (0, \frac{B_{-1}}{v}+\sqrt{\frac{B_{-1}}{B_{1}}}\right )
+ \nonumber\\
&& \Gamma (0, B_{-1}^{1/2}B_{1}^{3/2}) - \exp vB_{1}\Gamma (0, 
B_{-1}^{1/2}B_{1}^{3/2} + vB_{1}) +\nonumber\\
&& v(\exp\frac{B_{-1}}{v}\Gamma\left (0, \frac{B_{-1}}{v} + 
\sqrt{\frac{B_{-1}}{B_{1}}}\right ) - \Gamma\left (0,\sqrt{\frac{B_{-1}}
{B_{1}}}\right ))B_{1} )\\
&& - \sqrt{\frac{B_{-1}}{B_{1}}}v\exp [-B_{-1}^{1/2}B_{1}^{3/2}] +
B_{-1}(vB_{1}\Gamma (0, B_{-1}^{1/2}B_{1}^{3/2}) + \nonumber\\
&& \Gamma (0, B_{-1}^{1/2}B_{1}^{3/2}) - \exp [vB_1]\Gamma (0, 
B_{-1}^{1/2}B_{1}^{3/2}+vB_{1}))\}, \nonumber
\end{eqnarray}
where $v = M_{bc}^2 + M_{cc}^2$ and $\Gamma (a, z)$ is the incomplete
gamma function, which is given by the integral $\Gamma (a, z) =\int_z^{\infty}
t^{a-1}e^{-t}dt$.

\setlength{\unitlength}{1mm}

\end{document}